\documentclass[12pt,letterpaper]{article}
\pdfoutput=1
\usepackage{jheppub}
\usepackage{amsfonts,amsthm}
\usepackage[english]{babel}
\usepackage[utf8]{inputenc}
\usepackage{slashed}
\hypersetup{unicode}

\newcommand{\eq}{\begin{equation}}
\newcommand{\feq}{\end{equation}}
\newcommand{\eqn}{\begin{eqnarray}}
\newcommand{\feqn}{\end{eqnarray}}
\font\mybb=msbm10 at 12pt
\def\bb#1{\hbox{\mybb#1}}
\def\bZ {\bb{Z}}

\def\bR {\bb{R}}

\newcommand{\D}{{\rm d}}

\title{Rotating black holes in 4d gauged supergravity}

\author{Alessandra Gnecchi$^a$, Kiril Hristov$^b$, Dietmar Klemm$^{c}$, Chiara Toldo$^a$ and
Owen Vaughan$^d$}

\affiliation{$^a$ Institute for Theoretical Physics and Spinoza Institute, \\
\hspace*{0.15cm} Utrecht University, 3508 TD Utrecht, The Netherlands. \\
$^b$ Dipartimento di Fisica, Universit\`a di Milano-Bicocca, \\
\hspace*{0.15cm} Piazza della Scienza 3, 20126 Milano, Italy, and \\
\hspace*{0.15cm} INFN, sezione di Milano--Bicocca, Piazza della Scienza 3, 20126 Milano, Italy. \\
$^c$ Dipartimento di Fisica, Universit\`a di Milano, \\
\hspace*{0.15cm} Via Celoria 16, 20133 Milano, Italy, and \\
\hspace*{0.15cm} INFN, sezione di Milano, Via Celoria 16, 20133 Milano, Italy. \\
$^d$ Department of Mathematics and Center for Mathematical Physics, \\
\hspace*{0.15cm} University of Hamburg, Bundesstrasse 55, 20146 Hamburg, Germany.}

\preprint{ITP-UU-13/28 \\ \hspace*{\fill} SPIN-13/19 \\ \hspace*{\fill} IFUM-1018-FT \\ \hspace*{\fill} ZMP-HH/13-21}

\abstract{We present new results towards the construction of the most general black hole solutions in four-dimensional Fayet-Iliopoulos gauged supergravities. In these theories black holes can be asymptotically AdS
and have arbitrary mass, angular momentum, electric and magnetic charges and NUT charge. Furthermore, a wide range of horizon topologies is allowed (compact and noncompact) and the complex scalar fields have a nontrivial radial and angular profile. We construct a large class of solutions in the simplest single scalar model with
prepotential $F = -i X^0 X^1$ and discuss their thermodynamics. Moreover, various approaches
and calculational tools for facing this problem with more general prepotentials are presented.\vspace{0.5cm}}


\begin{document}
\maketitle
\flushbottom

\section{Introduction}
The construction and physical understanding of 4d black hole solutions in supersymmetric theories with negative cosmological constant is relevant for a number of developments in high energy physics. One can try to analyze such solutions in their own right as string theory ground states and understand black hole thermodynamics \cite{Bekenstein:1973ur,Hawking:1974sw,Hawking:1982dh,Hawking:1998kw,Chamblin:1999tk,Cvetic:1999ne,Chamblin:1999hg,Caldarelli:1999xj,Hristov:2013sya} and microscopic degeneracy \cite{Strominger:1996sh,Maldacena:1997de,Dijkgraaf:1996it,Strominger:1997eq}, guided by the AdS/CFT correspondence. In this sense black holes are the best test ground for the fundamental principles of quantum gravity, and therefore the knowledge of all possible black hole solutions in a given theory can be regarded as a first step in the programme of solving this theory on a quantum level. Alternatively, these gravitational systems provide non-trivial asymptotically AdS backgrounds with holographic duals that exhibit a rich structure and a wide range of applications in field theory and condensed matter systems, see
e.g.~\cite{Hartnoll:2009sz,Herzog:2009xv,McGreevy:2009xe,Sachdev:2010ch,Barisch:2011ui,Anninos:2013mfa}.

In the present paper we address the question of how to find generic AdS black hole solutions
in theories with $\text{U}(1)$ gauge fields and scalars that arise in the framework of gauged
supergravity. Such solutions will be labeled by a set of conserved charges, corresponding to the global symmetries of the system - in four dimensions these are the mass, NUT charge, angular momentum, electric and magnetic charges of the $\text{U}(1)$ gauge fields, and possible scalar hair. Unlike asymptotically flat black holes which
only have spherical horizons, solutions in AdS can be further distinguished by their horizon topology \cite{Vanzo:1997gw}. In the compact case the horizons can be Riemann surfaces
of arbitrary genus, while noncompact horizons correspond for example to black brane solutions. 

There have been numerous partial results on the topic in the last decade. Due to the lack of electromagnetic duality in the electrically gauged theories that are usually considered\footnote{Note that restoring the duality only
rotates electric and magnetic charges, without changing the types of black hole classes, thus the discussion
remains true, upto redefinition of the meaning of electric and magnetic charges.}, the various classes of solutions that were found look considerably different depending on the types of charges switched on. The known classes of electric solutions include both extremal and thermal, static and rotating solutions \cite{Duff:1999gh,Chong:2004na,Chow:2010sf,Chow:2010fw}, where in the BPS limit the known solutions are necessarily rotating \cite{Caldarelli:1998hg,AlonsoAlberca:2000cs} with constant scalars. The available magnetic black holes\footnote{Dyonic AdS black holes were constructed recently in \cite{Lu:2013ura}.}
can instead be BPS only in the static limit \cite{Cacciatori:2009iz} (with the exception of one hyperbolic rotating solution \cite{Klemm:2011xw}) with nonconstant scalars, see also \cite{Dall'Agata:2010gj,Hristov:2010ri}\footnote{See \cite{Halmagyi:2013sla,Halmagyi:2013qoa} for black hole solutions of this type in theories with nontrivial hypermultiplets.}. Their non-BPS and thermal generalizations were also known only with vanishing angular momentum \cite{Klemm:2012yg,Toldo:2012ec,Klemm:2012vm,Gnecchi:2012kb}. We show that all these seemingly disjoint classes in fact fall inside a single general solution, once we allow not only for rotation and electric and magnetic charges, but also for NUT charge. This extra freedom allows us to find a large parameter space of black hole solutions in one of the possible models (with prepotential $F=-iX^0 X^1$) and gives strong hints on how to tackle the same problem with more complicated scalar manifolds.

We also discuss an alternative but complementary approach to constructing solutions which is based on dimensional reduction and the real formulation of special geometry, as developed in \cite{Mohaupt:2011aa}. Within this formalism, the problem of constructing stationary solutions of $4D,\, {\cal N} = 2$ Fayet-Iliopoulos gauged supergravity reduces to solving a particular three-dimensional Euclidean non-linear sigma model (with potential). Previously this appraoch has been used to construct static black hole solutions \cite{Klemm:2012yg, Klemm:2012vm}, whereas in this paper we are interested in rotating solutions and the procedure is adapted accordingly. As an application, we present the new rotating black hole solutions of the $F=-iX^0 X^1$ model within this formalism, thus providing a useful consistency check.

The plan of the paper is as follows. We briefly present the Lagrangian and
equations of motion of the theory at the end of this section. Section \ref{universal} contains a general discussion of the universal
structure of black holes based on the Carter-Pleba\'nski metric \cite{Carter:1968ks,Plebanski:1975},
relating it to various examples existing in the literature. In the same section we comment on the difference between over- and under-rotating extremal solutions in AdS, and we explain how to choose different horizon topologies. In section \ref{prep1} we study the theory with prepotential $F=-iX^0 X^1$ and find a general class of nonextremal black holes with angular momentum and
magnetic charges. We discuss the thermodynamics and physical properties of these solutions, showing a new type of limit leading to noncompact horizons with finite area. In section \ref{nut}, these solutions
are extended to allow for NUT- and electric charges, however the NUT charge will not be allowed to take arbitrary
values. We also show how the solutions can be written in terms of harmonic functions and special geometry quantities and how in the limit of vanishing gauging a class of known solutions to ungauged supergravity
\cite{LozanoTellechea:1999my} is recovered.
We end the main discussion of the paper with some general comments and suggestions in section \ref{outlook}. Some of the useful tools and techniques we used to obtain our main results are relegated to the appendices. In app.~\ref{1/2BPS} we consider supersymmetric
rotating attractors, showing that all asymptotically flat under-rotating attractors are precisely half-BPS (in \cite{Hristov:2012nu} they were shown
to be at least quarter-BPS). In app.~\ref{AppB} we give more details about the real formulation of special geometry.

\vspace*{0.5cm}

{\bf Note added:} During the write-up of our work, ref.~\cite{Chow:2013gba} appeared, where charged rotating
solutions of the same model were presented.

\subsection{Lagrangian and equations of motion}
A detailed description of notations and conventions of abelian $\mathcal{N}=2$ gauged supergravity can be found in \cite{Andrianopoli:1996cm}. In the context of black hole physics, such models are also discussed in \cite{Cacciatori:2009iz,Klemm:2010mc,Dall'Agata:2010gj,Hristov:2010ri}, and we refer the reader to those papers for a complete introduction.

The most general bosonic Lagrangian of ${\cal N}=2$ abelian Fayet-Iliopoulos (FI) gauged
supergravity is given by\footnote{We use the convention $F_{\mu \nu}=\frac12 (\partial_{\mu} A_{\nu} -
\partial_{\nu} A_{\mu})$ for the field strengths. Our signature is mostly plus, and
$\epsilon_{0123}=-\epsilon^{0123}=1$.}
\begin{equation}\label{act}
e^{-1}{\cal L} = \frac12 R -g_{i \bar{\jmath}} \partial^{\mu}z^{i} \partial_{\mu} \bar{z}^{\bar{\jmath}} + I_{\Lambda \Sigma}(z, \bar{z}) F_{\mu \nu}^{\Lambda} F^{\mu \nu |\Sigma} +\frac12 R_{\Lambda \Sigma} (z, \bar{z}) \epsilon^{\mu \nu \rho \sigma} F_{\mu \nu}^{\Lambda} F_{\rho \sigma}^{\Sigma} - g^2 V(z, \bar{z})\,,
\end{equation}
with $\Lambda, \Sigma= 0,1... , n_V$ and $i,j=1,..., n_V$, where $n_V$ is the number of vector multiplets.
The imaginary and the real part of the period matrix $\mathcal{N}_{\Lambda \Sigma}$ ($I_{\Lambda \Sigma}$ and $R_{\Lambda \Sigma}$), as well as the metric on the scalar moduli space $g_{i \bar{\jmath}}$ and the scalar potential $V$, depend on the particular vector multiplet model. The complex scalars $z^i$ are written in terms of the holomorphic symplectic sections $(X^{\Lambda},F_{\Lambda})$. All these quantities can be specified uniquely just with a single holomorphic function, $F(X)$, the prepotential. Therefore specifying the prepotential is
equivalent to defining the full Lagrangian.

The Einstein equations following from \eqref{act} read
$$
-(R_{\mu \nu} - \frac12 g_{\mu \nu} R) = g_{\mu \nu} g^2 V(z, \bar{z}) + g_{\mu \nu} \partial^{\sigma}z^{i} \partial_{\sigma} \bar{z}^{\bar{\jmath}} g_{i \bar{\jmath}}-2  g_{i \bar{\jmath}} \partial_{(\mu}z^{i} \partial_{\nu)}\bar{z}^{\bar{\jmath}} +
$$
\begin{equation}
- I_{\Lambda \Sigma} g_{\mu \nu} F_{\rho \sigma}^{\Lambda} F^{\rho \sigma | \Sigma} +4 I_{\Lambda \Sigma}  F_{\mu \alpha}^{\Lambda} {F_{\nu}}^{\alpha|\Sigma}\,,
\end{equation}
while the equations of motion for the scalar fields $z^i$ are given by
\begin{equation}
g_{i \bar{\jmath}} \partial_{\mu}(e \partial^{\mu}\bar{z}^{\bar{\jmath}})+ e\frac{\partial g_{i \bar{k}}}{\partial \bar{z}^{\bar{\jmath}}} \partial^{\mu}\bar{z}^{\bar{\jmath}} \partial_{\mu} \bar{z}^{\bar{k}}+ e \frac{\partial I_{\Lambda \Sigma}}{\partial z^{i}} F_{\rho \sigma}^{\Lambda} F^{ \rho \sigma |\Sigma} + \frac{e}{2} \frac{\partial R_{\Lambda \Sigma}}{\partial z^{i}} \epsilon^{\mu \nu \rho \sigma} F_{\mu \nu}^{\Lambda} F_{\rho \sigma}^{\Sigma} - e g^2 \frac{\partial V}{\partial z^{i}}=0\,,
\end{equation}
and the Maxwell equations for the vector fields $A_{\nu}^{\Lambda}$ are
\begin{equation}
\partial_{\mu}(e F^{\mu \nu \,| \Sigma} I_{\Sigma \Lambda} +\frac{e}{2} \epsilon^{\mu\nu\rho\sigma}F_{\rho \sigma}^{\Sigma} R_{\Lambda \Sigma})=0\,.
\end{equation}
The requirement of an asymptotic $\text{AdS}_4$ geometry fixes the values of the scalar fields at infinity.
Indeed, $\text{AdS}_4$ corresponds to an extremum of the scalar potential, yielding the asymptotic
attractor condition 
\begin{eqnarray}
\left.\frac{\partial V}{\partial z^{i}}\right|_{\infty}=0\ , \qquad \Leftrightarrow \qquad
\xi_{\Lambda} D_iX^{\Lambda}\big|_{\infty}=0\ ,
\end{eqnarray}
where $D_i X^{\Lambda}$ is the K\"ahler covariant derivative of the coordinates $X^{\Lambda}$ with respect to~$z^i$. The constants $\xi_{\Lambda}$ determine which linear combination $\xi_{\Lambda}A^{\Lambda}$
is used to gauge a $\text{U}(1)$ subgroup of the $\text{SU}(2)$ R-symmetry. In what follows, we shall use
$g_{\Lambda}\equiv g\xi_{\Lambda}$, with $g$ the gauge coupling constant appearing in \eqref{act}.

\section{Universal structure of rotating black holes}
\label{universal}

This section is devoted to showing that all known rotating black holes in matter-coupled ${\cal N}=2$
gauged supergravity in four dimensions have a universal metric structure. It turns out that in all cases the
metric can be cast in the form
\eq
ds^2 = -f(dt + \omega_ydy)^2 + f^{-1}\left[v\left(\frac{dq^2}{Q} + \frac{dp^2}{P}\right) + PQ dy^2\right]\,,
\label{fibration}
\feq
where $Q(q)$ and $P(p)$ are polynomials of fourth degree respectively in the variables $q$ (radial variable) and $p$ (function of the angular variable $\theta$). The warp factors $f$, $v$, $\omega_y$ are more general functions of $q$ and $p$. We start now with the examples and finish the section by commenting on some novel general features of this metric, such as the difference between over- and under-rotating solutions and the relation between the function $P(p)$ and the choice of horizon topology.

\vspace{2mm}

\subsection{Carter-Pleba\'nski solution} 

The metric and $\text{U}(1)$ field strength of the Carter-Pleba\'nski solution \cite{Carter:1968ks,Plebanski:1975}
of minimal gauged supergravity are respectively given by
\eq
ds^2 = -\frac{Q(q)}{p^2+q^2}(d\tau - p^2d\sigma)^2 + \frac{p^2+q^2}{Q(q)}dq^2
                + \frac{p^2+q^2}{P(p)}dp^2 + \frac{P(p)}{p^2+q^2}(d\tau + q^2d\sigma)^2\,, 
                      \label{metr-CP}
\feq
\eq
F = \frac{\mathsf{Q}(p^2-q^2)+2\mathsf{P}pq}{(p^2+q^2)^2}\D q\wedge (\D\tau - p^2\D\sigma)
+ \frac{\mathsf{P}(p^2-q^2)-2\mathsf{Q}pq}{(p^2+q^2)^2}\D p\wedge (\D\tau + q^2\D\sigma)\,, \label{F-CP}
\feq
where the quartic structure functions read
\begin{eqnarray}
P(p) &=& \alpha - \mathsf{P}^2 + 2np - \varepsilon p^2  + (-\Lambda/3)p^4\,, \nonumber \\
Q(q) &=& \alpha + \mathsf{Q}^2 - 2mq + \varepsilon q^2 + (-\Lambda/3)q^4\,. \label{struc-func}
\end{eqnarray}
Here, $\mathsf{Q}$, $\mathsf{P}$ and $n$ denote the electric, magnetic and NUT-charge respectively, $m$
is the mass parameter, while $\alpha$ and $\varepsilon$ are additional non-dynamical
constants.

By making the coordinate transformation
\eq
\tau = At+By\,, \qquad \sigma = Ct+Dy\,, \qquad AD-BC=1\,,
\feq
\eqref{metr-CP} can be cast into the form \eqref{fibration},
where
\eq
v = Q(A-p^2C)^2 - P(A+q^2C)^2\,, \qquad f = \frac v{p^2+q^2}\,,
\feq
and
\eq
\omega_y = \frac1v\left[Q(A-p^2C)(B-p^2D)-P(A+q^2C)(B+q^2D)\right]\,.
\feq
We see that there is actually more than one way to write \eqref{metr-CP} as a fibration \eqref{fibration}
over a three-dimensional base space. A simple choice would be for instance $A=D=1$, $B=C=0$,
such that
\begin{displaymath}
v = Q - P\,, \qquad \omega_y = \frac{Qp^2+Pq^2}{P-Q}\,.
\end{displaymath}

\vspace{2mm}

\subsection{Rotating magnetic BPS black holes, prepotential $F=-iX^0X^1$}

Our second example is the family of BPS magnetic rotating black holes in the model with prepotential
$F=-iX^0X^1$, constructed in \cite{Klemm:2011xw}. 

This model has just one complex
scalar $\tau$. The symplectic sections in special coordinates are
$
v^T= \left( 1 , \tau , -i\tau , -i\right)
$. 
The K\"ahler potential, metric and  vector kinetic matrix are
respectively of this form:
\eq
e^{-{\cal K}} = 2(\tau + \bar\tau)\ , \qquad g_{\tau\bar\tau} = \partial_\tau\partial_{\bar\tau}
{\cal K} = (\tau + \bar\tau)^{-2}\ ,
\feq
\eq
{\cal N} = \left(\begin{array}{cc} -i\tau & 0 \\ 0 & -\frac i\tau\end{array}\right)\ ,
\feq
thus requiring
${\mathrm{Re}}\tau>0$. For our choice of electric gauging, the scalar potential is
\eq
V = -\frac4{\tau+\bar\tau}(g_0^2 + 2g_0g_1\tau + 2g_0g_1\bar\tau
+ g_1^2\tau\bar\tau)\ , \label{pot_su11}
\feq
which has an extremum at $\tau=\bar\tau=|g_0/g_1|$.

The metric of the BPS solution of \cite{Klemm:2011xw} reads
\begin{eqnarray}
ds^2 = && \frac{p^2+q^2-\Delta^2}{P}dp^2 + \frac{P}{p^2+q^2-\Delta^2}\left(dt + (q^2-\Delta^2)
dy\right)^2 \nonumber \\
&& + \frac{p^2+q^2-\Delta^2}{Q}dq^2 - \frac{Q}{p^2+q^2-\Delta^2}\left(dt - p^2dy\right)^2\ ,
\label{metr-X0X1-CP}
\end{eqnarray}
with the structure functions
\eq
P = (1+A)\frac{\mathsf{E}^2l^2}4 - \mathsf{E}p^2 + \frac{p^4}{l^2}\ , \qquad
Q = \frac1{l^2}\left(q^2 + \frac{\mathsf{E}l^2}2 - \Delta^2\right)^2\ . \label{struc-func-BPS}
\feq
The upper parts of the (nonholomorphic) symplectic section $(L^{\Lambda},M_{\Lambda})$ and the
$\text{U}(1)$ gauge potentials are given by
\begin{displaymath}
L^0 = \frac12\!\left(\frac{g_1}{g_0}\right)^{\frac12}\!\!\left(\frac{p^2+(q-\Delta)^2}{p^2+q^2-\Delta^2}
\right)^{\frac12}\,, \quad L^1 = \frac12\!\left(\frac{g_0}{g_1}\right)^{\frac12}\!\!\frac{p^2+q^2-\Delta^2+
2ip\Delta}{\left[(p^2+q^2-\Delta^2)(p^2+(q-\Delta)^2)\right]^{\frac12}}\,,
\end{displaymath}
\begin{displaymath}
A^{\Lambda} = -\frac{\mathsf{E}p\sqrt{-A}}{4g_{\Lambda}(p^2+q^2-\Delta^2)}(dt + (q^2-\Delta^2)dy)\,,
\qquad \Lambda=0,1\,.
\end{displaymath}
The solution is thus specified by three free parameters $A,\mathsf{E},\Delta$. (The asymptotic AdS
curvature radius $l$ is related to the gauge coupling constants by $l^{-2}=4g_0g_1$). The new rotating
solution that we are going to describe in section \ref{prep1} is a nonextremal deformation of this solution.
For $\Delta=0$, the moduli are constant, and the solution reduces to a subclass
of \eqref{metr-CP}, \eqref{F-CP}.

The metric \eqref{metr-X0X1-CP} can again be written in the form \eqref{fibration}, where now
\eq
v = Q-P\,, \qquad f = \frac v{p^2+q^2-\Delta^2}\,,
\feq
\eq
\omega_y = \frac{P(q^2-\Delta^2)+Qp^2}{P-Q}\,.
\feq
\subsection{Rotating black holes of Chong, Cveti\v{c}, Lu, Pope and Chow}

Finally, there are three other examples of rotating black hole solutions, described in \cite{Chong:2004na,Chow:2010sf,Chow:2010fw}. They all fit in the form of the metric \eqref{fibration}.
We report the details of the rotating black holes with two pair-wise equal charges in $\text{SO}(4)$-gauged
${\cal N}=4$ supergravity constructed in \cite{Chong:2004na}, since they are the most relevant for the
new configurations described in section \ref{prep1}. The metric, dilaton, axion and gauge
fields read respectively
\begin{eqnarray}
ds^2 &=& -\frac{\Delta_r}W(dt - a\sin^2\!\theta d\phi)^2 + W\left(\frac{dr^2}{\Delta_r} + \frac{d\theta^2}
                  {\Delta_{\theta}}\right) + \frac{\Delta_{\theta}\sin^2\!\theta}W\left[adt - (r_1r_2 + a^2)d\phi
                  \right]^2\,, \nonumber \\
e^{\varphi_1} &=& \frac{r_1^2 + a^2\cos^2\!\theta}W = 1 + \frac{r_1(r_1-r_2)}W\,, \qquad
                                  \chi_1 = \frac{a(r_2-r_1)\cos\theta}{r_1^2+a^2\cos^2\!\theta}\,, \nonumber \\
A^1 &=& \frac{2\sqrt2 ms_1c_1\left[adt - (r_1r_2 + a^2)d\phi\right]\cos\theta}W\,, \nonumber \\
A^2 &=& \frac{2\sqrt2 m s_2c_2r_1(dt-a\sin^2\!\theta d\phi)}W\,, \label{sol-cvetic}
\end{eqnarray}
where
\begin{eqnarray}
\Delta_r &=& r^2 + a^2 - 2mr + g^2r_1r_2(r_1r_2 + a^2)\,, \qquad
                        \Delta_{\theta} = 1 - a^2g^2\cos^2\!\theta\,, \\
W &=& r_1r_2 + a^2\cos^2\!\theta\,, \qquad r_I = r + 2ms_I^2\,, \qquad s_I = \sinh\delta_I\,, \qquad c_I =
             \cosh\delta_I\,. \nonumber
\end{eqnarray}
Notice that the other scalar fields $\varphi_2,\varphi_3,\chi_2,\chi_3$ are set to zero in the truncation
of \cite{Chong:2004na}. Also, the two electromagnetic charges of the solution are carried by fields
in $\text{U}(1)$ subgroups of the two $\text{SU}(2)$ factors in $\text{SO}(4)\sim\text{SU}(2)\times\text{SU}(2)$.
In the case $\delta_1=\delta_2$, the dilaton $\varphi_1$ and the axion $\chi_1$ vanish. Then, the
solution boils down to the Kerr-Newman-AdS geometry with purely electric charge if $A^1$ is dualized,
or to purely magnetic KNAdS if we dualize $A^2$. 
Note also that, after dualizing $A^1$, the model considered in \cite{Chong:2004na} (their Lagrangian (52))
can be embedded into ${\cal N}=2$ gauged supergravity as well, by choosing the prepotential
$F=-iX^0X^1$ \cite{Colleoni:2012jq}.

After the rescaling $\phi \to ay$ and the redefinition $q=r$, $p=a\cos\theta$, the metric in
\eqref{sol-cvetic} can again be cast into the form \eqref{fibration}, with (quartic) structure
functions\footnote{It should be emphasized that, like in the case of the Carter-Pleba\'nski solution, also
here and in the previous example there is an $\text{SL}(2,\bR)$ gauge freedom that consists in
sending $t\mapsto\alpha t+\beta y$, $y\mapsto\gamma t+\delta y$, $\alpha\delta-\beta\gamma=\pm1$,
which preserves the form \eqref{fibration} of the metric while transforming the functions $f$, $v$ and
$\omega_y$. This freedom can prove useful for the explicit construction of new solutions.
\label{foot-sl2R}}
\eq
P = (1 - g^2p^2)(a^2 - p^2)\,, \qquad Q = \Delta_r\,,
\feq
and
\eq
v = Q-P\,, \qquad f = \frac vW\,,
\feq
\eq
\omega_y =  \frac{Q(p^2-a^2)+P(r_1r_2+a^2)}{v}\,.
\feq

As mentioned before, the solutions of \cite{Chow:2010sf,Chow:2010fw} can be recast in the form \eqref{fibration} too. The reader can find the complete form of the solution in the original papers, here we skip the details, since the procedure is straightforward and along the same lines as for the previous ones.

We first rescale
$ t \to \Xi t $, $\phi \to ay $,
and redefine $q=r$, $p=a\cos\theta$. Then, for the single-charged black hole solution of \cite{Chow:2010sf} the structure functions are:
\begin{equation}\label{PandQ}
P = (1 - g^2p^2)(a^2 - p^2)\,, \qquad Q = \Delta_r\,,
\end{equation}

\begin{equation}\label{formofV}
v = \left((1 - g^2p^2)Q - V_r^2(a^2 - p^2)\right)(1 - g^2p^2)\,,
\end{equation}

\begin{equation}
 f = \frac v{(H)^{\frac12}(p^2+q^2)}\,, \qquad \omega_y = \frac{2m q c P \sqrt{1+a^2g^2s^2}}{v\Xi}\,,
\end{equation}
while for the the two-charge rotating black holes of \cite{Chow:2010fw} the functions $P$, $Q$ and $v$ have the same form as \eqref{PandQ} \eqref{formofV}, whereas
\eq
 \qquad f = \frac v{(H_1H_2)^{\frac12}(p^2+q^2)}\,, \qquad 
\omega_y = \frac{2mrc_1c_2\tilde c_1\tilde c_2P}{v\Xi}\,.
\feq
The simplicity of the geometry \eqref{fibration}, and the fact that it is particularly suited for a formalism
based on timelike dimensional reduction like the one used in \cite{Mohaupt:2011aa}, should help constructing
new nonextremal rotating black holes in matter-coupled gauged supergravity with an arbitrary number of
vector multiplets and general prepotentials. Unfortunately even if the universal structure should remain the
same, the equations of motion of gauged supergravity depend crucially on the given model and cannot be
solved in complete generality, therefore we first restrict ourselves to considering the simplest interesting
prepotentials with a single vector multiplet.

\subsection{Over- vs.~under-rotating solutions}

An interesting possibility arises in the extremal limit of rotating black holes (see
e.g.~\cite{Rasheed:1995zv,Astefanesei:2006dd}). One can sometimes find several extremal limits that
correspond to either of two physically different solutions, called over-rotating and under-rotating solutions.
The over-rotating solutions (a typical example here is the extremal Kerr black hole) have an ergoregion, while
the under-rotating (that resemble more the extremal Reissner-Nordstr\"{o}m spacetime) do not have
ergoregions. Due to the AdS asymptotics, allowing for a wide range of coordinate choices, it might not be easy to see immediately whether one can have both types of extremal limits. The key to determining this is the following - one first needs to write the black hole metric in asymptotically AdS coordinates, from which the asymptotic time direction can be extracted. Once we know the correct Killing vector $k = \partial_t$, we can follow its behavior on the horizon. For an under-rotating solution, $k$ is null, $|k|^2=0$, while an overrotating solution has $|k|^2 > 0$, indicating the existence of an ergosphere.

To make the discussion more explicit, let us take the example of our metric \eqref{fibration},
\eq
ds^2 = -f(dt + \omega_y dy)^2 + f^{-1}\left[v\left(\frac{dq^2}{Q} + \frac{dp^2}{P}\right) + PQ dy^2\right]\,,
\feq
and assume for the sake of argument that this metric was already written in asymptotically AdS coordinates\footnote{Note that this in general will not be true for the explicit solutions we find and one needs to first perform a coordinate change from the Pleba\'nski form of the metric to the asymptotically AdS form.}
(this means that in the limit $q\rightarrow\infty$, one has $\omega_y = 0, f \sim q^2, v \sim q^4, Q \sim q^4$).
In the extremal limit, with horizon at $q_h$, $Q(q_h) = Q'(q_h) = 0$, the norm of the timelike Killing vector is
$-f$ and $f(q_h)$ will be either vanishing or negative. Typically both these possibilities will exist for some
choice of parameters that determine the solution. This leads to three distinct physical possibilities for the
complete geometry:
\begin{itemize}

\item $f(q_h) < 0$, only possible if $\omega_y (q_h) \neq 0$: this corresponds to the over-rotating solution; typically there is a lower bound for the angular momentum, $|J| > J_{\text{min}}$ (sometimes
$J_{\text{min}} = 0$).

\item $f(q_h) = 0$ and $\omega_y (q_h) = 0$: static attractor, leading to a static black hole, typically resulting from the limit $J = 0$.

\item $f(q_h) = 0$, $\omega_y (q_h) \neq 0$: under-rotating solution; typically there is an upper bound
for the angular momentum, $0< |J| < J_{\text{max}}$. 

\end{itemize}
The first two cases exist as extremal limits for all known rotating solutions with electromagnetic charges, while the third case is quite special and exists only in the presence of nontrivial scalar fields. Under-rotating solutions are known to exist in ungauged supergravity with cubic prepotentials, \cite{Rasheed:1995zv,Astefanesei:2006dd}, but not for quadratic ones of the type $F=-iX^0X^1$. We were not able to find explicit examples of under-rotating solutions in AdS among the general solutions of the $F=-iX^0X^1$ discussed in the present paper, but their existence in other models is an interesting possibility. 

Note that it is important to identify the asymptotic time to be able to properly distinguish between the two rotating cases, thus only an analysis of the near-horizon geometry is in principle not enough, even if it gives good hints of the nature of the solution. In particular, observe that the near-horizon geometries of the asymptotically flat under- and over-rotating solutions are exactly the same (cf.~(5.38) and (5.60) of \cite{Astefanesei:2006dd}), but the
former are defined in the parameter space $J^2<\mathsf{P}^2\mathsf{Q}^2$, while the latter only for
$J^2>\mathsf{P}^2\mathsf{Q}^2$, where $\mathsf{P}$ and $\mathsf{Q}$ denote the magnetic and electric
charge respectively.

\subsection{Relation between $P(p)$ and horizon topology}
\label{Hor_topology}
The horizon topology of the black hole with metric \eqref{fibration} can be studied through the analysis of the function $P(p)$.
It is a quartic polynomial in $p$ and here we choose to write it in the form 
\begin{equation}
	P(p) = (p-p_a) (p-p_b) (p-p_c) (p-p_d)\ ,
\end{equation}
where the roots $p_{a,b,c,d}$ depend on the explicit values of the physical parameters of the metric (mass, NUT charge, electric and magnetic charges). 

If we first look at the simple case without NUT charge, we have pairs of roots such that $p_c = - p_b, p_d = -p_b$. In order for the induced metric on the horizon $q=q_{\text h}$ (where $Q(q_{\text h})=0$) to have the right signature, we need $P\ge 0$. Let us first assume that the polynomial $P$ has four real roots with $0<p_a<p_b$. Then $P$ is non-negative for $|p|\le p_a$ or $|p|\ge p_b$. Choosing $-p_a \leq p \leq p_a$ leads to a function $P(p)$ that is bounded and vanishes at two points, which are coordinate singularities. Such a function can be defined for horizons with spherical topology, where the two singularities correspond to the north and the south pole of the (possibly squashed) sphere. Choosing the other possibility, $p_b \leq p$, leads to hyperbolic topology since the function $P$ is not bounded anymore. The coordinate singularity at $p = p_b$ is at the origin of the hyperbolic space in the standard hyperbolic coordinates. One can see that in this case we have two disjoint types of black holes within the same solutions, depending on whether we choose the compact or non-compact range. The third main type of topology arises in the case when $p_a$ and $p_b$ are both complex, thus $P$ is everywhere positive and non-vanishing for real $p$ - this corresponds to the flat topology of black branes, where no coordinate singularities are encountered. 

To summarize the three basic types of topology and their relation with $P(p)$, the possibilities are
\begin{itemize}
	\item spherical topology: $P(p)$ bounded and vanishing at two points, north and south pole.
	
	\item hyperbolic topology: $P(p)$ unbounded and vanishing at a single point.
	
	\item flat topology: $P(p)$ unbounded and never vanishing.
\end{itemize}

On top of those topologies and their quotients, we can have some new exotic situations in some special cases. In section \ref{noncomp-hor} we will show the situation where the two positive roots of $P$ coincide, $p_a = p_b$. It turns out choosing $-p_a \leq p \leq p_a$ in this case leads to a sphere with two punctures on the place of the two poles, i.e.\ the horizon has a cylindrical topology but finite area. Thus we are lead to think that whenever the function $P(p)$ has a double root the horizon is punctured at that point, which is no longer just a coordinate singularity. 

The situation with NUT charge is even more complex, since then all four roots can be a priori unrelated to each other. One can therefore have situations with $p_d<p_c<p_b = p_a$ for example, where the choice of bounded region for $P(p)$ will lead to one pole and one puncture and therefore to a bottle-shaped horizon topology. Even more exotic possibilities would be three coinciding roots, a case which is yet to be analyzed carefully. In any case, the three main types of horizon topologies continue to exist whether one allows for NUT charge or not.

\section{Thermal rotating solutions with magnetic charges}
\label{prep1}

We shall now construct a nonextremal deformation of the BPS solution to the model with prepotential
$F=-iX^0X^1$, constructed in \cite{Klemm:2011xw}, and described in the previous section. Inspired by the
form (58) of \cite{Chong:2004na}, we can make the ansatz
\eq
ds^2 = -\frac QW\left[dt-p^2dy\right]^2 + \frac PW\left[dt + q_1q_2dy\right]^2 + W\left(\frac{dq^2}Q
             + \frac{dp^2}P\right)\,, \label{ansatz-nonextr}
\feq 
with
\eq
Q = a_0 + a_1q + a_2q^2 + a_4q^4\,, \qquad P = b_0 + b_1p + b_2p^2 + b_4p^4\,, \label{QP}
\feq
and
\eq
W = q_1q_2 + p^2\,, \qquad q_i = q - \Delta_i\,,
\feq
where $a_i$, $b_i$, and $\Delta_i$ are constants. \eqref{ansatz-nonextr} fits into the general form of
the metric \eqref{fibration} with
\eq
v = Q-P\,, \qquad f = \frac vW\,, \qquad \omega_y = - \frac1v\left[ P q_1 q_2 + Q p^2 \right]\,.
\feq
It boils down to the BPS solution \eqref{metr-X0X1-CP} when $Q,P$ reduce to the functions
\eqref{struc-func-BPS}, and $\Delta_1=-\Delta_2\equiv\Delta$.
The ansatz for the gauge potentials and the scalars is
\eq
A^\Lambda = \frac{\mathsf{P}^\Lambda(dt + q_1q_2dy)}Wp\,, \qquad \tau = e^{-\varphi} + i\chi =
\frac{X^1}{X^0} = \frac{\mu W+i\nu p}{q_1^2+p^2}\,, \label{ansatz-A-tau-magn}
\feq
where the constants $\mathsf{P}^\Lambda$ are proportional to the magnetic charges, and $\mu,\nu$ are
real constants.

In order to reduce the number of free parameters, we will first restrict to the case
$\Delta_1=-\Delta_2\equiv\Delta$, and take $\mu,\nu$ in the scalar to be the
same as in the BPS case. We have then checked that the equations of motion are satisfied if
\begin{equation}
a_0 = b_0 - a_2 \Delta^2 - \frac{\Delta^4}{l^2} + 2l^2\left(g_0^2{\mathsf{P}^0}^2 +
g_1^2{\mathsf{P}^1}^2\right)\,, \qquad a_1= \frac{2 l^2 (g_0^2{\mathsf{P}^0}^2 - g_1^2{\mathsf{P}^1}^2)}
{ \Delta}\,, \label{a0a1}
\end{equation}
\begin{equation}
b_1=0\,, \qquad  b_2 = -a_2 - 2 \frac{\Delta^2}{l^{2}}\,, \qquad  b_4 = a_4 = 1/l^2 \equiv 4g_0g_1\,,
\end{equation}
\begin{equation}\label{nu}
\mu=\frac{g_0}{g_1}\,, \qquad \nu= 2\frac{\Delta g_0}{g_1}\,,
\end{equation}
where we assumed that $g_0,g_1$ are positive. We can check that the scalar field $\tau$ has the correct
behaviour at infinity, since in this model the $\text{AdS}_4$ asymptotic geometry is obtained for
$\tau_{\infty}=g_0/g_1$. 
If we fix the Fayet-Iliopuolos constants $g_0$ and $g_1$, the solution depends on the five parameters $b_0$,
$a_2$, $\Delta$, $\mathsf{P}^0$, $\mathsf{P}^1$, thus two more parameters with respect to the BPS solution. 
From the second equation of \eqref{a0a1} we see that, in the case $g_0\mathsf{P}^0=g_1\mathsf{P}^1$,
one has either $a_1=0$ or $\Delta=0$. If $a_1$ vanishes, one can thus have equal charges and yet
a nontrivial scalar profile (i.e., $\Delta\neq 0$). This behaviour is qualitatively different from that of
the solutions constructed for instance in \cite{Chong:2004na}.

Notice that the form of the scalar field and of the vector field strengths is the same as in the BPS case.
The latter is recovered for
\begin{equation}
a_2=\mathsf{E} - 2 \frac{ \Delta^2}{l^2}\,, \qquad b_0 = (1 + A)\frac{\mathsf{E}^2 l^2}4\,, \qquad
\mathsf{P}^{\Lambda}=\frac{\sqrt{-A}\mathsf{E}}{4g_{\Lambda}}\,.
\end{equation}

A further generalization to a black hole with electric and NUT-charges (that would include also the solutions
of \cite{Klemm:2011xw} and \cite{Chong:2004na}) is
straightforward, however, we postpone this discussion to the next section and first elaborate on the physical properties and novelties of the magnetic solutions. In this case, in fact, the absence of closed timelike curves
makes them interesting thermodynamical and gravitational systems.

\subsection{Physical discussion}
\label{phys-disc}

Following section \ref{Hor_topology}, we assume $P$ has four distinct roots, $\pm p_a$, $\pm p_b$, where $0<p_a<p_b$. Then $P$ is non-negative for $|p|\le p_a$ or $|p|\ge p_b$. Since we are interested in black holes with compact horizon\footnote{\label{footnote-hyperb}As already discussed, noncompact
hyperbolic horizons can be obtained by restricting to the region $p\ge p_b$ and setting
$p=p_b\cosh\theta$, where $0\le\theta<\infty$. In this case, the rotation parameter is defined by
$p_b^2=j^2$. The resulting black holes represent generalizations of the solutions of minimal gauged
supergravity constructed in \cite{Klemm:1997ea}.}, we consider the range $|p|\le p_a$,
and set $p=p_a\cos\theta$, where $0\le\theta\le\pi$. By using the scaling symmetry
\begin{equation}
p \to \lambda p\,, \qquad q \to \lambda q\,, \qquad t \to t/\lambda\,, \qquad y \to y/\lambda^3\,, \qquad
\Delta \to \lambda \Delta\,, \label{scaling-symm}
\end{equation}
\begin{displaymath}
a_0 \to \lambda^4 a_0\,, \qquad a_1 \to \lambda^3 a_1\,, \qquad
a_2 \to \lambda^2 a_2\,, \qquad b_0 \to \lambda^4 b_0\,, \qquad b_2 \to \lambda^2 b_2\,, 
\end{displaymath}
one can set $p_b=l$ without loss of generality. If we define the rotation parameter $j$ by
$p_a^2=j^2$, this amounts to the choice
\eq
b_0 = j^2\,, \qquad b_2 = -1-\frac{j^2}{l^2}\,,
\feq
which implies
\begin{equation}
a_0 = (j^2-\Delta^2)\left(1-\frac{\Delta^2}{l^2}\right) + 2l^2\left(g_0^2{\mathsf{P}^0}^2 +
g_1^2{\mathsf{P}^1}^2\right)\,, \qquad a_2 = 1-\frac{\Delta^2}{l^2} + \frac{j^2-\Delta^2}{l^2}\,. \label{a0a2}
\end{equation}
Taking also
\eq
t \to t + \frac{j\phi}{\Xi}\,, \qquad y \to \frac{\phi}{j\Xi}\,, \qquad \Xi \equiv 1 - \frac{j^2}{l^2}\,,
\label{shift-t}
\feq
the metric \eqref{ansatz-nonextr} becomes
\begin{eqnarray}
ds^2 = && -\frac{Q}{(q^2-\Delta^2+j^2\cos^2\!\theta)}\left[dt + \frac{j\sin^2\!\theta}{\Xi} d\phi\right]^2 +
                   (q^2-\Delta^2+j^2\cos^2\!\theta)\left(\frac{dq^2}{Q} + \frac{d\theta^2}{\Delta_{\theta}}\right) \nonumber \\
             && +\, \frac{\Delta_{\theta}\sin^2\!\theta}{(q^2-\Delta^2+j^2\cos^2\!\theta)}\left[jdt +
                   \frac{q^2+j^2-\Delta^2}{\Xi}d\phi\right]^2\,, \label{metr-nonextr-spher-X0X1}
\end{eqnarray}
where we defined
\begin{displaymath}
\Delta_{\theta} = 1 - \frac{j^2}{l^2}\cos^2\!\theta\,.
\end{displaymath}
From \eqref{a0a1} it is clear that for $g_0\mathsf{P}^0=g_1\mathsf{P}^1$ and $\Delta=0$, the mass
parameter $a_1$ can be arbitrary; this leads to the Kerr-Newman-AdS solution with magnetic charge
and constant scalar. On the other hand, for zero rotation parameter, $j=0$,
\eqref{metr-nonextr-spher-X0X1} boils down to the static nonextremal black holes with
running scalar constructed in \cite{Klemm:2012yg}.

The Bekenstein-Hawking entropy of the black holes described by \eqref{metr-nonextr-spher-X0X1}
is given by
\eq\label{entropy}
S = \frac{\pi}{\Xi G}(q_{\text h}^2 + j^2 - \Delta^2)\,,
\feq
where $G$ denotes Newton's constant and $q_{\text h}$ is the location of the horizon, i.e.,
$Q(q_{\text h})=0$. In order to compute the temperature and angular velocity, we write the metric
in the canonical (ADM) form
\eq
ds^2 = -N^2dt^2 + \sigma(d\phi - \omega dt)^2 + (q^2-\Delta^2+j^2\cos^2\!\theta)\left(\frac{dq^2}{Q} +
\frac{d\theta^2}{\Delta_{\theta}}\right)\,,
\feq
with
\eq
\sigma = \frac{\Sigma^2\sin^2\!\theta}{(q^2-\Delta^2+j^2\cos^2\!\theta)\Xi^2}\,, \qquad
\omega = \frac{j\Xi}{\Sigma^2}(Q-\Delta_{\theta}(q^2+j^2-\Delta^2))\,,
\feq
and the lapse function
\eq
N^2 = \frac{Q\Delta_{\theta}(q^2-\Delta^2+j^2\cos^2\!\theta)}{\Sigma^2}\,,
\feq
where
\begin{displaymath}
\Sigma^2 \equiv \Delta_{\theta}(q^2+j^2-\Delta^2)^2-Qj^2\sin^2\!\theta\,.
\end{displaymath}
The angular velocity of the horizon is thus
\eq
\omega_{\text h} = \omega|_{q=q_{\text h}} = -\frac{j\Xi}{q_{\text h}^2 + j^2 - \Delta^2}\,, \label{omega_h}
\feq
whereas at infinity one has
\eq
\omega_{\infty} = \frac j{l^2}\,. \label{omega_inf}
\feq
The angular momentum computed by means of the Komar integral reads
\begin{equation}
J = \frac1{16\pi G}\oint_{\text{S}^2_{\infty}} dS^{\mu\nu}\nabla_{\mu}m_{\nu}\,, \label{J-Komar}
\end{equation}
with $m=\partial_{\phi}$ and the oriented measure
\begin{displaymath}
dS^{\mu\nu} = (v^{\mu}u^{\nu} - v^{\nu}u^{\mu})\sqrt{\hat\sigma}d\theta d\phi\,.
\end{displaymath}
Here, $u=N^{-1}(\partial_t+\omega\partial_{\phi})$ is the normal vector of a constant $t$
hypersurface, $v=(Q/(q^2-\Delta^2+j^2\cos^2\theta))^{1/2}\partial_q$, and
\begin{displaymath}
\sqrt{\hat\sigma} = \frac{\Sigma\sin\theta}{\Xi\Delta_{\theta}^{1/2}}\,,
\end{displaymath}
where $\hat\sigma$ denotes the induced metric on a two-sphere of constant $q$ and $t$.
Evaluation of \eqref{J-Komar} yields
\begin{equation}\label{J}
J = \frac{a_1j}{2\Xi^2G}\,.
\end{equation}
The Komar mass
\eq
M = -\frac1{8\pi G}\oint_{\text{S}^2_{\infty}} dS^{\mu\nu}\nabla_{\mu}k_{\nu}
\feq
has to be computed with respect to the Killing vector $k=\Xi^{-1}\partial_t$ \cite{Caldarelli:1999xj},
leading to
\begin{displaymath}
M = -\frac1{8\pi G}\lim_{q\to\infty}\int d\theta d\phi\frac{\sin\theta}{(j^2-l^2)^2}\left[-2l^2q^3 - 2(j^2-\Delta^2)l^2q
+ l^4a_1 + {\cal O}(q^{-1})\right]\,,
\end{displaymath}
which is of course divergent. If we subtract the background with $a_1=0$ and the same $j$ and
$\Delta$, we get the finite result
\eq
M = -\frac{a_1}{2\Xi^2G}\,. \label{KomarM}
\feq
Notice that the `ground state' with $a_1=0$ is a naked singularity (contrary to the case of hyperbolic
horizons addressed in footnote \ref{footnote-hyperb}): The curvature singularity $W=0$\footnote{Note
also that for $W<0$, the real part of the scalar field becomes negative, so that ghost modes
appear.} is shielded by a horizon if $q_{\text h}^2-\Delta^2+j^2\cos^2\!\theta>0$, which is equivalent to
\begin{displaymath}
\frac2{l^2}q_{\text h}^2 + a_2 > 1 + \frac{j^2}{l^2}\,.
\end{displaymath}
Now, using $Q(q_{\text h})=0$, this can be rewritten as
\begin{displaymath}
\sqrt{a_2^2-\frac{4a_0}{l^2}} > 1 + \frac{j^2}{l^2}\,,
\end{displaymath}
which can be easily shown to lead to a contradiction by using \eqref{a0a2}.

An alternative mass definition, that does not require any background subtraction, is based on the
Ashtekar-Magnon-Das (AMD) formalism \cite{Ashtekar:1984zz,Ashtekar:1999jx}.
(Cf.~also \cite{Chow:2010sf} for an application to rotating AdS black holes and for more details).
First of all we compute the Weyl
tensor of a conformally rescaled metric (in this case the conformal rescaling factor is $\Omega=
l/q$), to leading order in $q$. This reads
\begin{equation}
\overline{C}^{t}_{qtq} = \frac{-g_0^2 {\mathsf{P}^0}^2 + g_1^2 {{\mathsf{P}^1}}^2}{8 \Delta g_0^2 g_1^2 q^5} +O(1/q^6)\,.
\end{equation}
Once we have this quantity, we can compute the mass associated to the Killing vector
$K = \Xi^{-1}\partial_t$, given by
\begin{equation}
M= \frac{1}{8 \pi G (4 g_0 g_1)^{3/2}}\int_{\Sigma} d\overline{\Sigma}_a \Omega^{-1} \bar{n}^c \bar{n}^d \overline{C}_{cbd}^a K^b= -\frac{(g_0^2 {{\mathsf{P}^0}}^2 - g_1^2 {{\mathsf{P}^1}}^2)}{4\Delta g_0 g_1 \,G \,\Xi^2} = -\frac{a_1}{2 \,G \, \Xi^2}\,,
\end{equation}
so that the AMD procedure gives the same result as the
regularized Komar integral.

The magnetic charges $\pi^\Lambda$ are given by
\eq
\pi^\Lambda = \frac1{4\pi}\oint_{\text{S}^2_{\infty}} F^\Lambda = -\frac{\mathsf{P}^\Lambda}{\Xi}\,.
\feq
Now that we have computed the physical quantities of our solution, a comment on the number of free
parameters is in order. We already mentioned that the metric \eqref{ansatz-nonextr} and the gauge potentials
and scalar \eqref{ansatz-A-tau-magn} depend on five parameters. However, due to the scaling
symmetry \eqref{scaling-symm}, one of them is actually redundant and can be scaled away. There remain
thus four free parameters, for instance $P^\Lambda,\Delta,j$, or alternatively $\pi^\Lambda,M,J$. Our
black holes are therefore labelled by two indpendent magnetic charges, mass and angular momentum.
Note also that the parameter $\Delta$ related to the running of the scalar is not independent of the mass;
for this reason our solution does not carry primary hair. 

The product of the horizon areas, given formula \eqref{entropy}, is
\eq
\prod_{\alpha=1}^4 A_{\alpha} = \frac{(4\pi)^4}{\Xi^4} \prod_{\alpha=1}^4 (q_{\text h_{\alpha}}^2 + j^2 - \Delta^2) =  \frac{(4\pi)^4}{\Xi^4} \prod_{\alpha=1}^4 (q_{\text h_{\alpha}}- q_{+}) (q_{\text h_{\alpha}} -q_{-})\,,
\feq 
with $q_{\pm} = \pm \sqrt{\Delta^2-j^2}$. At this point the formulas resemble the ones given in the static case, and we can use the procedure explained in \cite{Toldo:2012ec}. We define
\eq
\kappa_+= q_{+}^4 + \frac{a_2}{a_4} q_{+}^2 + \frac{a_1}{a_4} q_{+} +\frac{a_0}{a_4} \qquad
\kappa_-= q_{-}^4 + \frac{a_2}{a_4} q_{-}^2 + \frac{a_1}{a_4} q_{-} +\frac{a_0}{a_4} \,,
\feq
so that the area product will be given by $ \prod_{\alpha=1}^4 A_{\alpha}= (4\pi)^4\kappa_+\kappa_-/\Xi^4$.
Plugging in the values of the coefficients and using the expression \eqref{J} for $J$ we have
\eq
 \prod_{\alpha=1}^4 A_{\alpha} = (4\pi)^4 l^2\left((\pi^0\pi^1)^2+J^2\right)\,. \label{area-prod}
\feq
The charge-dependent term on the rhs of \eqref{area-prod} is directly related to the prepotential; a fact that
was first noticed in \cite{Toldo:2012ec} for static black holes.

\subsection{Thermodynamics and extremality}
\label{thermodynamics}
A quasi-Euclidean section of the metric can be obtained by analytically continuing $t\to -it_{\text E}$.
It turns out that this is regular at $q=q_{\text h}$ provided $t_{\text E}$ is identified modulo
$4\pi\Xi(q_{\text h}^2+j^2-\Delta^2)/Q_{\text h}'$, where $Q_{\text h}'$ denotes the derivative of $Q$
w.r.t.~$q$, evaluated at the horizon. This yields the Hawking temperature
\eq
T = \frac{Q_{\text h}'}{4\pi(q_{\text h}^2+j^2-\Delta^2)}\,. \label{temp}
\feq
Using the expressions \eqref{entropy}, \eqref{J} and \eqref{KomarM} for the entropy,
angular momentum and mass respectively, as well as the fact that $Q$ vanishes for $q=q_{\text h}$,
one obtains by simple algebraic manipulations the Christodoulou-Ruffini-type mass formula
\begin{eqnarray}
M^2 &=& \frac{S}{4\pi G} + \frac{\pi J^2}{SG} + \frac{\pi}{4SG^3}(\pi^0\pi^1)^2 + \left(\frac{l^2}{G^2} +
                 \frac{S}{\pi G}\right)\left((g_0\pi^0)^2 + (g_1\pi^1)^2\right) \nonumber \\
         & & + \frac{J^2}{l^2} + \frac{S^2}{2\pi^2l^2} + \frac{S^3G}{4\pi^3l^4}\,. \label{CR}
\end{eqnarray}
Note that this reduces correctly to equ.~(43) of \cite{Caldarelli:1999xj} in the KNAdS case
$(g_0\pi^0)^2=(g_1\pi^1)^2$, $\Delta=0$, $a_1$ arbitrary.

Since $S,J,\pi^\Lambda$ form a complete set of extensive parameters, \eqref{CR} represents also the black hole thermodynamic fundamental relation $M=M(S,J,\pi^\Lambda)$. The quantities conjugate to $S,J,\pi^\Lambda$
are the temperature
\begin{eqnarray}
T = \left(\frac{\partial M}{\partial S}\right)_{\!J,\pi^\Lambda} = \frac1{8\pi GM}&&\left[1 - \frac{4\pi^2J^2}{S^2}
           - \frac{\pi^2}{S^2G^2}(\pi^0\pi^1)^2 + 4\left((g_0\pi^0)^2 + (g_1\pi^1)^2\right)\right. \nonumber \\
    & & \,\left. + \frac{4SG}{\pi l^2} + \frac{3S^2G^2}{\pi^2l^4}\right]\,, \label{T}
\end{eqnarray}
the angular velocity
\eq
\Omega = \left(\frac{\partial M}{\partial J}\right)_{\!S,\pi^\Lambda} = \frac{\pi J}{MGS}\left[1 + \frac{SG}{\pi l^2}\right]\,,
\label{Omega}
\feq
and the magnetic potentials
\eq
\Phi_\Lambda = \left(\frac{\partial M}{\partial\pi^\Lambda}\right)_{\!S,J,\pi^{\Sigma\neq\Lambda}} =
\frac1{MG}\left[\frac{\pi}{4SG^2}\pi^0\pi^1\eta_{\Lambda\Sigma}\pi^\Sigma + \left(\frac{l^2}{G} +
\frac{S}{\pi}\right)g_\Lambda^2\pi^\Lambda\right]\,, \label{magn-pot}
\feq
where
\begin{displaymath}
\eta_{\Lambda\Sigma} = \left(\begin{array}{cc} 0 & 1 \\ 1 & 0 \end{array}\right)\,,
\end{displaymath}
and there is no summation over $\Lambda$ in the last term. The obtained quantities satisfy the first law
of thermodynamics
\eq
dM = TdS + \Omega dJ + \Phi_\Lambda d\pi^\Lambda\,.
\feq
Furthermore, by eliminating $M$ from \eqref{T}-\eqref{magn-pot} using \eqref{CR}, it is possible
to obtain four equations of state for the black holes \eqref{metr-nonextr-spher-X0X1}.
It is straightforward to verify that the relation \eqref{T} for the temperature coincides with equ.~\eqref{temp},
whereas \eqref{Omega} yields
\eq
\Omega = \omega_{\text h} - \omega_{\infty}\,,
\feq
with $\omega_{\text h}$ and $\omega_{\infty}$ given respectively by \eqref{omega_h} and
\eqref{omega_inf}. It is thus the difference between the angular velocities at the horizon and
at infinity which enters the first law; a fact that was stressed in \cite{Caldarelli:1999xj} for the case
of the KNAdS black hole.

The Hawking temperature \eqref{temp} vanishes in the extremal case, when $q_{\text h}$ is at least a
double root of $Q$. The structure function $Q$ can then be written as
\begin{displaymath}
Q = (q-q_{\text h})^2\left(\frac{q^2}{l^2} + \frac{2q_{\text h}q}{l^2} + a_2 + \frac{3q_{\text h}^2}{l^2}\right)\,,
\end{displaymath}
and we must have
\eq
a_0 = a_2q_{\text h}^2 + \frac{3q_{\text h}^4}{l^2}\,, \qquad a_1 = -2q_{\text h}a_2 - \frac{4q_{\text h}^3}{l^2}\,.
\feq
These equations restrict of course the number of free parameters compared to the nonextremal case.
To obtain the near-horizon geometry of the extremal black holes, we define new (dimensionless)
coordinates $z,\hat t,\hat\phi$ by
\eq
q = q_{\text h} + \epsilon q_0 z\,, \qquad t = \frac{\hat t q_0}{\Xi\epsilon}\,, \qquad \phi = \hat\phi +
\omega_{\text h}\frac{\hat t q_0}{\epsilon}\,, \qquad q_0^2 \equiv \frac{\Xi l^2(q_{\text h}^2+j^2-\Delta^2)}
{6q_{\text h}^2+a_2l^2}\,, \label{zoom}
\feq
and take $\epsilon\to 0$ keeping $z,\hat t,\hat\phi$ fixed. This leads to
\begin{eqnarray}
ds^2 = && \frac{q_{\text h}^2-\Delta^2+j^2\cos^2\!\theta}{C}\left(-z^2 d\hat t^2 + \frac{dz^2}{z^2}
                   + C\frac{d\theta^2}{\Delta_{\theta}}\right) \nonumber \\
             &&\quad+\, \frac{\Delta_{\theta}\sin^2\!\theta(q_{\text h}^2+j^2-\Delta^2)^2}{(q_{\text h}^2 - \Delta^2
                   + j^2\cos^2\!\theta)}\left(\frac{d\hat\phi}{\Xi} + \frac{2q_{\text h}\omega_{\text h}}C z d\hat t\right)^2\,,
\end{eqnarray}
where the constant $C$ is given by
\begin{eqnarray}
C &=& \frac{6q_{\text h}^2}{l^2} + a_2 \nonumber \\
    &=& \left\{\left(1-\frac{\Delta^2}{l^2}\right)^2 + \frac{(j^2-\Delta^2)^2}{l^4} + \frac{14}{l^2}\left(1 -
             \frac{\Delta^2}{l^2}\right)(j^2 - \Delta^2) + 24(g_0^2{\mathsf{P}^0}^2 + g_1^2{\mathsf{P}^1}^2)
             \right\}^{1/2}\,. \nonumber
\end{eqnarray}
If $q_{\text h}$ is at least a triple root of $Q$, $C$ vanishes, and one has an ultracold black hole.
In this case, the zooming procedure \eqref{zoom} does not conform to Geroch's criteria of limiting
spaces \cite{Geroch:1969ca}, and thus the resulting geometry would not even solve the equations of motion.
This problem was first pointed out by Romans \cite{Romans:1991nq}, and discussed also in
\cite{Meessen:2010ph}. There exists an alternative limiting procedure \cite{Ginsparg:1982rs,Cardoso:2004uz}
which basically consists in going first to the situation where $Q$ has a double root, and then taking the
near-horizon limit simultaneously with the ultracold limit in a particular way. We postpone a discussion
of the ultracold case to a future publication.

Note that in the extremal limit, when $T=0$, it is easy to see that the entropy is only a function of the discrete charges $J$ and $\pi^I$ by inverting \eqref{T} in terms of $S$.

\subsection{Noncompact horizon with finite area}
\label{noncomp-hor}

We shall now discuss the special case where the polynomial $P(p)$ has two double roots, i.e.,
$p_a=p_b$ in the notation adopted at the beginning of section \ref{phys-disc}. This corresponds
to $j^2=l^2$, which means that the conformal boundary rotates at the speed of light. For the
Kerr-AdS solution, this limit (in which the metric \eqref{metr-nonextr-spher-X0X1} is of course singular)
was explored in \cite{Hawking:1998kw}\footnote{See also \cite{Caldarelli:2008pz,Caldarelli:2012cm}.},
where it was argued that it represents an interesting example in which to study AdS/CFT.

Using again the scaling symmetry \eqref{scaling-symm}, we can set $p_a=l$ without loss of generality,
so that
\begin{displaymath}
P(p) = \frac1{l^2}(p^2 - l^2)^2
\end{displaymath}
in this case. The induced metric on the horizon $q=q_{\text h}$ (where $Q$ vanishes) is given by
\eq
ds_{\text h}^2 = \frac{P}{q_{\text h}^2 - \Delta^2 + p^2}(q_{\text h}^2 - \Delta^2 + \alpha)^2 dy^2 +
\frac{q_{\text h}^2 - \Delta^2 + p^2}{P}dp^2\,, \label{metr-hor-double-root}
\feq
where the constant $\alpha$ takes into account a possible shift $t\to t+\alpha y$, similar to
\eqref{shift-t}. If we want $y$ to be a compact coordinate, the absence of closed timelike curves
requires setting $\alpha=l^2$, since otherwise $g_{yy}$ will be negative close to $p^2=l^2$.
Note that we consider the coordinate range $-l\le p\le l$, and that
\eqref{metr-hor-double-root} becomes singular for $p^2=l^2$. To understand more in detail what happens
at these singularities, take for instance the limit $p\to l$, in which \eqref{metr-hor-double-root}
simplifies to
\eq
ds_{\text h}^2 = (q_{\text h}^2 - \Delta^2 + l^2)\left[\frac{d\rho^2}{4\rho^2} + 4\rho^2 dy^2\right]\,.
\label{metr-hor-H2}
\feq
Here, the new coordinate $\rho$ is defined by $\rho=l-p$.
\eqref{metr-hor-H2} is clearly a metric of constant negative curvature on the hyperbolic
space ${\text H}^2$ (or on a quotient thereof, if we want $y$ to be a compact coordinate). Since
\eqref{metr-hor-double-root} is symmetric under $p\to -p$, an identical result holds for $p\to -l$.
Thus, for $p\to\pm l$, the horizon approaches a space of constant negative curvature, and there is
no true singularity there. In particular, this implies that the horizon is noncompact, which comes
as a surprise, since one might have expected the limit of coincident roots $p_a=p_b$ to be smooth,
and for $p_a\neq p_b$ the horizon was topologically a sphere. Moreover, the horizon area reads
\eq
A_{\text h} = \int (q_{\text h}^2 - \Delta^2 + l^2)dy dp = 2Ll(q_{\text h}^2 - \Delta^2 + l^2)\,,
\feq
where we assumed $y$ to be identified modulo $L$. We see that, in spite of being noncompact, the event horizon
has finite area, and the entropy of the corresponding black hole is thus also finite. To the best of our
knowledge, this represents the first instance of a black hole with noncompact horizon, but still finite
entropy.

\begin{figure}[htb]
  \begin{center}
    \includegraphics[scale=0.7]{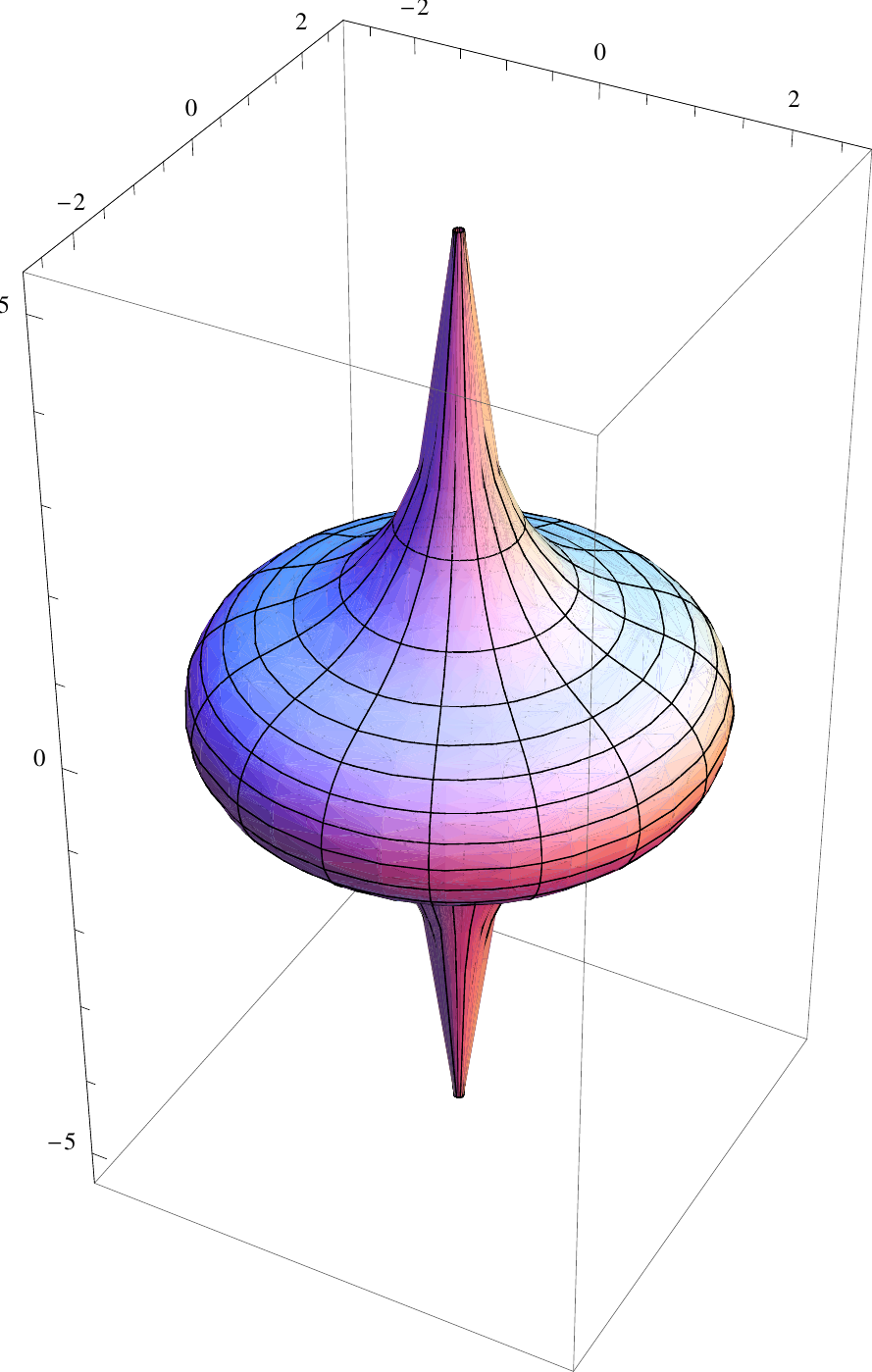}
  \end{center}
  \caption{The event horizon of a black hole in the case where $P(p)$ has two double roots, embedded in
  $\bR^3$ as a surface of revolution.\label{cusp}}
\end{figure}

In order to visualize the geometry \eqref{metr-hor-double-root}, we can embed it in $\bR^3$ as a surface of
revolution\footnote{This is possible if $L(q_{\text h}^2 - \Delta + l^2)$ is not too large, since otherwise
$(dz/dp)^2$ in \eqref{dz/dp} will become negative in some region.}. To this end write the flat metric
in cylindrical coordinates,
\begin{displaymath}
ds_3^2 = dz^2 + dr^2 + r^2 d\phi^2\,,
\end{displaymath}
and consider $z=z(p)$, $r=r(p)$. Setting $\phi=2\pi y/L$, and identifying the resulting line element with
\eqref{metr-hor-double-root}, one gets
\eq
r = \frac{L}{2\pi l}(q_{\text h}^2 - \Delta^2 + p^2)^{-1/2}(l^2 - p^2)(q_{\text h}^2 - \Delta^2 + l^2)\,,
\feq
as well as
\eq
\left(\frac{dr}{dp}\right)^2 + \left(\frac{dz}{dp}\right)^2 = \frac{q_{\text h}^2 - \Delta^2 + p^2}{P}\,, \label{dz/dp}
\feq
which is a differential equation for $dz/dp$. By expanding near $p=\pm l$, one easily sees that $z$ diverges
logarithmically for $|p|\to l$, and that $r$ goes to zero in this limit. We integrated \eqref{dz/dp} numerically
for the values $l=1$, $L=2\pi$ and $q_{\text h}^2-\Delta^2=5$. The resulting surface of
revolution is shown in figure \ref{cusp}, where the $z$-axis is vertical. Note that the two cusps extend up to infinity,
with $z\to\pm\infty$ for $p\to\pm l$ respectively, while the `equator' $z=0$, where $r$ becomes maximal,
is reached for $p=0$.

The metric on the conformal boundary $q\to\infty$ of the black hole solution reads
\eq
ds_{\text{bdry}}^2 = -dt^2 + 2dtdy(p^2-l^2) + l^2\frac{dp^2}{P}\,, 
\feq
and hence $y$ becomes a lightlike coordinate there.

\section{Inclusion of NUT- and electric charges}
\label{nut}

Inspired by the solution in section 5 of \cite{Chong:2004na}, we
make the following ansatz to include also NUT- and electric charges:
\eq
ds^2 = -\frac QW\left[dt-p_1p_2dy\right]^2 + \frac PW\left[dt + q_1q_2dy\right]^2 + W\left(\frac{dq^2}Q
             + \frac{dp^2}P\right)\,, \label{ansatz-nonextr-el}
\feq
where $Q,P$ are again given by \eqref{QP} (with $a_4=b_4=1/l^2$), and
\begin{displaymath}
W = q_1q_2 + p_1p_2\,, \qquad q_1 = q -\Delta\,, \qquad q_2 = q + \Delta\,, \qquad p_1 = p - \delta\,,
\qquad p_2 = p + \delta\,.
\end{displaymath}
The ansatz for the gauge potentials and the scalars is
\eq\label{Ai}
A^\Lambda = \frac{\mathsf{P}^\Lambda(dt + q_1q_2dy)}Wp_1 + \frac{\mathsf{Q}^\Lambda
(dt - p_1p_2dy)}Wq_1\,,
\feq
\eq
\tau = e^{-\varphi} + i\chi = \frac{X^1}{X^0} = \frac{\mu W+i(\nu p + \lambda q)}{q_1^2+p_1^2}\,,
\label{tau-NUT}
\feq
where the constants $\mathsf{Q}^\Lambda$ are proportional to the electric charges, and
$\mu,\nu,\lambda$ are constants to be determined.

We have checked that the equations of motion of the $F=-iX^0X^1$ model are satisfied if the parameters
assume the following form:
\eq
a_4=b_4=1/l^2\,, \label{eq:nut-a4}
\feq
\eq{\nonumber}
a_0= b_0 -a_2 \Delta^2 - \frac{\Delta^4}{ l^2} +\frac{g_0 {\mathsf{P}^0}^2}{2 g_1}+ \frac{g_1 {\mathsf{P}^1}^2}{2g_0}+\frac{g_0 {\mathsf{Q}^0}^2}{2 g_1}+ \frac{g_1 {\mathsf{Q}^1}^2}{2g_0}+
\feq
\eq
-\frac{a_2 \Delta^2 {\mathsf{Q}^1}^2}{{\mathsf{P}^1}^2}-\frac{2 \Delta^4 {\mathsf{Q}^1}^2}{ l^2 \, {\mathsf{P}^1}^2}-\frac{ \Delta^4 {\mathsf{Q}^1}^4}{l^2 \, {\mathsf{P}^1}^4}\,, \label{eq:nut-a0}
\feq
\eq
b_1= - \frac{ 2 l^2 \mathsf{P}^1 (g_0^2 (-2 \mathsf{P}^0 \mathsf{P}^1 \mathsf{Q}^0 + {\mathsf{P}^0}^2 \mathsf{Q}^1 - {\mathsf{Q}^0}^2 \mathsf{Q}^1) + 
    g_1^2 \mathsf{Q}^1 ({\mathsf{P}^1}^2 + {\mathsf{Q}^1}^2))}{\Delta ({\mathsf{P}^1}^2 + {\mathsf{Q}^1}^2)} \,,
    \label{eq:nut-b1}
\feq
\eq
a_1= - \frac{ 2 l^2 \mathsf{P}^1 (g_0^2 (-{\mathsf{P}^0}^2 \mathsf{P}^1 + \mathsf{P}^1 {\mathsf{Q}^0}^2 - 2 \mathsf{P}^0 \mathsf{Q}^0 \mathsf{Q}^1) + g_1^2 \mathsf{P}^1 ({\mathsf{P}^1}^2 +{\mathsf{Q}^1}^2))}{ \Delta ({\mathsf{P}^1}^2 + {\mathsf{Q}^1}^2)}\,,
\label{eq:nut-a1}
\feq
\eq
b_2=-a_2 - \frac{2 \Delta^2 ({\mathsf{P}^1}^2 + {\mathsf{Q}^1}^2)}{ l^2 \, {\mathsf{P}^1}^2}\,, \label{eq:nut-b2}
\feq
\eq
\mu=\frac{g_0}{g_1}\,, \qquad \nu=\frac{2 \Delta g_0}{g_1}\,, \qquad \lambda=\frac{2 \Delta g_0 \mathsf{Q}^1}{g_1 \mathsf{P}^1}\,, \qquad \delta=-\frac{\Delta \mathsf{Q}^1}{\mathsf{P}^1}\,. \label{eq:nut-mu}
\feq
The solution has free parameters $\Delta$, $b_0$ $a_2$, $\mathsf{P}^{\Lambda}$ and $\mathsf{Q}^{\Lambda}$. It reduces to the one we have previously found for $\mathsf{Q}^1=0=\mathsf{Q}^0$.

Let us now keep the ansatz \eqref{ansatz-nonextr-el}-\eqref{Ai} for the metric and the gauge fields and look for a solution with a scalar field of the form
\eq
\tau = e^{-\varphi} + i\chi = \frac{X^1}{X^0} = \frac{\mu W+i(\nu p + \lambda q)}{q_2^2+p_2^2}\,.
\label{tau-NUT-prime}
\feq
In this case, the equations of motions are satisfied for the following parameters
\eq
a_4=b_4=1/l^2\,,
\feq
\eq{\nonumber}
a_0= b_0 -a_2 \Delta^2 -\frac{ \Delta^4}{l^2} +\frac{g_0 {\mathsf{P}^0}^2}{2 g_1}+ \frac{g_1 {\mathsf{P}^1}^2}{2g_0}+\frac{g_0 {\mathsf{Q}^0}^2}{2 g_1}+ \frac{g_1 {\mathsf{Q}^1}^2}{2g_0}+
\feq
\eq
-\frac{a_2 \Delta^2 {\mathsf{Q}^0}^2}{{\mathsf{P}^0}^2}-\frac{2 \Delta^4  {\mathsf{Q}^0}^2}{ l^2 \, {\mathsf{P}^0}^2}-\frac{ \Delta^4 {\mathsf{Q}^0}^4}{l^2 \, {\mathsf{P}^0}^4}\,, \label{eq:nut-a0-prime}
\feq
\eq
b_1= - \frac{ 2 l^2 \mathsf{P}^0 (g_1^2 (-2 \mathsf{P}^1 \mathsf{P}^0 \mathsf{Q}^1 + {\mathsf{P}^1}^2 \mathsf{Q}^0 - {\mathsf{Q}^1}^2 \mathsf{Q}^0) + 
    g_0^2 \mathsf{Q}^0 ({\mathsf{P}^0}^2 + {\mathsf{Q}^0}^2))}{ \Delta  ({\mathsf{P}^0}^2 + {\mathsf{Q}^0}^2)} \,,
    \label{eq:nut-b1-prime}
\feq
\eq
a_1= - \frac{2 l^2 \mathsf{P}^0 (g_1^2 (-{\mathsf{P}^1}^2 \mathsf{P}^0 + \mathsf{P}^0 {\mathsf{Q}^1}^2 - 2 \mathsf{P}^1 \mathsf{Q}^0 \mathsf{Q}^1) + g_0^2 \mathsf{P}^0 ({\mathsf{P}^0}^2 +{\mathsf{Q}^0}^2))}{ \Delta ({\mathsf{P}^0}^2 + {\mathsf{Q}^0}^2)}\,,
\feq
\eq
b_2=-a_2 - \frac{2 \Delta^2  ({\mathsf{P}^0}^2 + {\mathsf{Q}^0}^2)}{ l^2 \,{\mathsf{P}^0}^2}\,,
\feq
\eq
\mu=\frac{g_0}{g_1}\,, \qquad \nu=-\frac{2 \Delta g_0}{g_1}\,, \qquad \lambda=-\frac{2 \Delta g_0 \mathsf{Q}^0}{g_1 \mathsf{P}^0}\,, \qquad \delta=-\frac{\Delta \mathsf{Q}^0}{\mathsf{P}^0}\,.
\label{eq:nut-mu-prime}
\feq
One can easily check that the two ans\"atze \eqref{tau-NUT} and \eqref{tau-NUT-prime} (when expressed
in terms of the parameters $\Delta$ and $\delta$) are related by the strong-weak coupling
transformation
\eq
\tau\to \frac{g_0^2}{g_1^2\tau}\,. \label{discrete-symm}
\feq
This is actually a residual $\bZ_2$ symmetry of the full symplectic group $\text{Sp}(4,\bR)$ that remains after the gauging, corresponding to a reparametrization of the scalar manifold, given by the action of the matrix
\eq
{\cal S} = \left(\begin{array}{cc} A & B \\ C & D\end{array}\right) \in \text{Sp}(4,\bR)\,, \label{sympl-transf}
\feq
with $B=C=0$,
\eq
A = \left(\begin{array}{cc} 0 & g_1/g_0 \\ g_0/g_1 & 0\end{array}\right)\,,
\feq
and $D=(A^{-1})^T$. Since ${\cal S}^2=\bb{I}$, it generates $\bZ_2$, as stated. Notice also that the scalar potential is invariant under \eqref{discrete-symm}. Moreover, the matrix \eqref{sympl-transf} acts on the charges by interchanging
$g_0\mathsf{P}^0\leftrightarrow g_1\mathsf{P}^1$ and $g_0\mathsf{Q}^0\leftrightarrow g_1\mathsf{Q}^1$,
which is exactly what transforms \eqref{eq:nut-a0}-\eqref{eq:nut-mu} to the new parameters 
\eqref{eq:nut-a0-prime}-\eqref{eq:nut-mu-prime}.

In order to discuss more in detail the solution \eqref{eq:nut-a4}-\eqref{eq:nut-mu}, we assume that the
polynomial $P$ has four distinct roots $p_a<p_b<p_c<p_d$. Since we are interested in black
holes with compact horizon\footnote{Black holes with hyperbolic horizons can be obtained by taking
the region $p\ge p_d$.},
we consider the region $p_b\le p\le p_c$ (where $P$ is positive), and
set $p=N+j\cos\theta$, where
\eq
N \equiv \frac{p_b+p_c}2\,, \qquad j \equiv \frac{p_c-p_b}2\,,
\feq
and $0\le\theta\le\pi$. Using the scaling symmetry \eqref{scaling-symm}, supplemented by
\begin{displaymath}
\delta \to \lambda\delta\,, \qquad b_1 \to \lambda^3 b_1\,,
\end{displaymath}
one can set $p_d=-N+\sqrt{l^2+4N^2}$ without loss of generality\footnote{This choice is made
in order to correctly reproduce the KNTN-AdS solution of minimal gauged supergravity as a
special subcase, see e.g.~\cite{AlonsoAlberca:2000cs}.}. This implies
\begin{displaymath}
b_0 = (j^2 - N^2)\left(1 + \frac{3N^2}{l^2}\right)\,, \quad b_1 = 2N\left(1 - \frac{j^2}{l^2} + \frac{4N^2}{l^2}
\right)\,, \quad b_2 = -1 - \frac{j^2}{l^2} - \frac{6N^2}{l^2}\,,
\end{displaymath}
\begin{displaymath}
a_0 = b_0 + b_2(\Delta^2 + \delta^2) + \frac{(\Delta^2 + \delta^2)^2}{l^2} + 2l^2\left[g_0^2({\mathsf{P}^0}^2
+ {\mathsf{Q}^0}^2) + g_1^2({\mathsf{P}^1}^2 + {\mathsf{Q}^1}^2)\right]\,,
\end{displaymath}
\begin{equation}
a_2 = 1 + \frac{j^2}{l^2} + \frac{6N^2}{l^2} - \frac2{l^2}(\Delta^2 + \delta^2)\,. \label{eq:nut-a2}
\end{equation}
Taking also
\begin{displaymath}
t \to t + \left(\frac{j}{\Xi} + \frac{N^2-\delta^2}{j\Xi}\right)\phi\,, \qquad y \to \frac{\phi}{j\Xi}\,,
\end{displaymath}
the metric \eqref{ansatz-nonextr-el} becomes
\begin{eqnarray}
ds^2 = && -\frac{Q}{q^2-\Delta^2+(N+j\cos\theta)^2-\delta^2}\left[dt + \frac{j\sin^2\!\theta}{\Xi} d\phi
                    -\frac{2N}{\Xi}\cos\theta d\phi\right]^2 \nonumber \\
             &&  +\, \left[q^2-\Delta^2+(N+j\cos\theta)^2-\delta^2\right]\left(\frac{dq^2}{Q} + \frac{d\theta^2}
                    {\Delta_{\theta}}\right) \label{metr-nonextr-spher-X0X1-NUT} \\
             && +\, \frac{\Delta_{\theta}\sin^2\!\theta}{q^2-\Delta^2+(N+j\cos\theta)^2-\delta^2}\left[jdt +
                   \frac{q^2+j^2+N^2-\Delta^2-\delta^2}{\Xi}d\phi\right]^2\,, \nonumber
\end{eqnarray}
where now
\begin{displaymath}
\Delta_{\theta} = 1-\frac{j^2}{l^2}\cos^2\!\theta-\frac{4Nj}{l^2}\cos\theta\,,
\end{displaymath}
while the fluxes and the scalar field are given respectively by
\begin{eqnarray}
F^\Lambda &=& \frac{\mathsf{P}^\Lambda(q^2-\Delta^2-(p-\delta)^2)-2(q-\Delta)p\mathsf{Q}^\Lambda}{\left[q^2-\Delta^2+(N+j\cos\theta)^2
              -\delta^2\right]^2}\sin\theta\left[jdt + \frac{q^2+j^2+N^2-\Delta^2-\delta^2}{\Xi}d\phi\right]
              \wedge d\theta \nonumber \\
     &+& \frac{\mathsf{Q^\Lambda}(p^2-\delta^2-(q-\Delta)^2)-2(p-\delta)q\mathsf{P}^\Lambda}{\left[q^2-\Delta^2+(N+j\cos\theta)^2
             -\delta^2\right]^2}dq\wedge\left[dt + \frac{j\sin^2\!\theta}{\Xi} d\phi -\frac{2N}{\Xi}\cos\theta d\phi\right]\,,
             \nonumber
\end{eqnarray}
\begin{displaymath}
\tau = \frac{g_0}{g_1}\frac{q+\Delta-i(p+\delta)}{q-\Delta-i(p-\delta)}\,,
\end{displaymath}
with $p=N+j\cos\theta$.

If one turns off the rotation ($j=0$), and fixes the charges in terms of $N,\Delta,\delta$ according to
\begin{displaymath}
g_1\mathsf{P}^1 = \frac{N^2}{l^2} - \frac{N\delta}{l^2} + \frac14\,, \quad g_1\mathsf{Q}^1 = -\frac{\delta}{\Delta}
g_1\mathsf{P}^1\,, \quad g_0\mathsf{P}^0 = g_1\mathsf{P}^1 +\frac{2N\delta}{l^2}\,, \quad g_0\mathsf{Q}^0 =
g_1\mathsf{Q}^1 + \frac{2N\Delta}{l^2}\,,
\end{displaymath}
one recovers the spherical NUT-charged BPS solution constructed in \cite{Colleoni:2012jq}\footnote{The flat
or hyperbolic BPS solutions of \cite{Colleoni:2012jq} can be obtained in a similar way. Notice that only the latter
represent genuine black holes, while in the spherical or flat case one has naked singularities \cite{Colleoni:2012jq}.}.
With the charges fixed as above, and
\begin{displaymath}
a_1 = -\frac{4N}{\Delta}\left[\frac{2N\delta^2}{l^2} + \frac{2N\Delta^2}{l^2} - \frac{2N^2\delta}{l^2} - \frac{\delta}2
         \right]\,,
\end{displaymath}
all the constraints \eqref{eq:nut-a0}-\eqref{eq:nut-b2} are satisfied.

From \eqref{metr-nonextr-spher-X0X1-NUT}, we can also get a dyonic solution without NUT charge.
Setting $N=0$ one has
\begin{displaymath}
b_0 = j^2\,, \qquad b_1 = 0\,, \qquad b_2 = -1 - \frac{j^2}{l^2}\,, \qquad a_2 = 1 + \frac{j^2}{l^2} - \frac2{l^2}
(\Delta^2 + \delta^2)\,,
\end{displaymath}
\begin{displaymath}
a_0 = j^2 - \left(1 + \frac{j^2}{l^2}\right)(\Delta^2 + \delta^2) + \frac{(\Delta^2 + \delta^2)^2}{l^2} + 2l^2\left[
          g_0^2({\mathsf{P}^0}^2 + {\mathsf{Q}^0}^2) + g_1^2({\mathsf{P}^1}^2 + {\mathsf{Q}^1}^2)\right]\,,
\end{displaymath}
and $a_1$ is given by \eqref{eq:nut-a1}. Since $b_1$ vanishes, \eqref{eq:nut-b1} implies $\mathsf{P}^1=0$ or
\begin{equation}
g_0^2\left(\mathsf{Q}^0\mathsf{Q}^1 + \mathsf{P}^0\mathsf{P}^1\right)^2 = \left({\mathsf{Q}^1}^2
+ {\mathsf{P}^1}^2\right)\left(g_1^2{\mathsf{Q}^1}^2 + g_0^2{\mathsf{P}^0}^2\right)\,,
\end{equation}
which allows to express e.g.~$\mathsf{Q}^0$ in terms of the other charges. The solution is thus specified by
the five parameters $\mathsf{P}^0,\mathsf{P}^1,\mathsf{Q}^1,j,\Delta$, or alternatively by three charges,
angular momentum and mass.

Note that, also in the case with nonvanishing $N$, \eqref{eq:nut-b1} together with the second eqn.~in
\eqref{eq:nut-a2} fix one of the electromagnetic charges in terms of the other parameters, and therefore
the solution is labelled by three independent $\text{U}(1)$ charges, NUT charge, angular momentum
and mass. It is thus not the most general solution, which should have four independent $\text{U}(1)$
charges.

\subsection{Solution with harmonic functions and flat limit}
We can partially rewrite the ansatz in terms of complex harmonic functions in order to make the dependence on the prepotential more suggestive. If we define the variable $\rho = q - i p$, we can use the harmonic functions in $\rho$,
\begin{equation}
	X^0 \equiv H_0 = h_0 (1 - \frac{\Delta - i \delta}{\rho})\ , \qquad X^1 \equiv H_1 = h_1 (1 + \frac{\Delta - i \delta}{\rho})\ ,
	\label{sections_ansatz}
	\end{equation}
with $h_0 = g_0^{-1}, h_1 = g_1^{-1}$. We can then rewrite the scalar ansatz as
\begin{equation}
	\tau = \frac{H_1}{H_0}\,,
\end{equation} 
while the function $W$ appearing in the metric and gauge field ans\"atze \eqref{ansatz-nonextr-el},
\eqref{Ai} can be cast into the form
\eq
	W = l^2 (q^2 + p^2) e^{-K (X^{\Lambda})}\ ,
\feq
where $K$ is the K\"ahler potential of special geometry that depends on the prepotential\footnote{Note that here we do not mean the physical K\"ahler potential $K(\tau, \bar{\tau})$, but the one that is obtained directly from the sections \eqref{sections_ansatz}.}. In the case of $F = -i X^0 X^1$, we have
\eq
	e^{-K}_{F = -i X^0 X^1} = \frac{1}{l^2 (q^2 + p^2)} (q_1 q_2 + p_1 p_2)\ ,
\feq
as needed. 

Rewriting the solution in this form makes it easy to take the limit of vanishing gauging.
We take $g_0, g_1 \rightarrow 0$, keeping the ratio an arbitrary finite constant (which is the value of the scalar field at infinity). This leads to a simplification in the explicit parameters $a_i, b_i$ that parametrize the functions $P(p)$ and $Q(q)$. We can again write the metric in the form \eqref{metr-nonextr-spher-X0X1-NUT}, but now with $\Delta_{\theta} = \Xi = 1$. A further redefinition of the radial coordinate $q = r + m$ for $a_1 = -2 m$ leads to $$Q = r^2 + (j^2+m^2-N^2-\Delta^2-\delta^2)\ .$$
Written this way, the solution can be seen to sit inside the general class of solutions of \cite{LozanoTellechea:1999my} with arbitrary mass, angular momentum, electric and magnetic and NUT charges. Just like in the case with cosmological constant, we cannot recover the most general class due to the restriction that the NUT charge is fixed in terms of the electric and magnetic charges, c.f.\ \eqref{eq:nut-b1} and \eqref{eq:nut-a2}.

\section{Final remarks and outlook}
\label{outlook}
Given the various examples and solutions we presented in the preceeding sections, we can make an ansatz that is likely to yield solutions for more general prepotentials with arbitrary number of vector multiplets. The metric ansatz would remain 
\eq
ds^2 = -f(dt + \omega_ydy)^2 + f^{-1}\left[v\left(\frac{dq^2}{Q} + \frac{dp^2}{P}\right) + PQ dy^2\right]\,,
\feq
with
\eq
Q = a_0 + a_1q + a_2q^2 + g^2 q^4\,, \quad P = b_0 + b_1p + b_2p^2 + g^2 p^4\,, \quad v = Q - P\ , \quad f = v e^{2 U}\ , 
\feq 
\eq
 e^{-2 U} = i (  \overline{X}^\Lambda F_\Lambda- X^\Lambda \overline{F}_\Lambda ) \ , \quad
\omega_y = - \frac{1}{v} \left(Q (c_0 + c_1 p + p^2) + P (d_0+d_1 q + q^2)\right)\ ,
\feq
scalar fields given by the symplectic sections
\eq
	X^\Lambda = h^\Lambda (q-i p + \Delta^\Lambda - i \delta^\Lambda)\ ,
\feq
and gauge fields
\begin{eqnarray}
A^\Lambda &=& \frac{1}{W} \left(\mathsf{P}^\Lambda (p+k_0) ({\rm d}t + (d_0+d_1 q + q^2) {\rm d} y)\right.
\nonumber \\
&&+\left. \mathsf{Q}^\Lambda (q+l_0) ({\rm d} t - (c_0 + c_1 p + p^2) {\rm d} y)\right)\ .
\end{eqnarray}
The real constant parameters $a_0, a_1, a_2, b_0, b_1, b_2, c_0, c_1, d_0. d_1, k_0, l_0, h^\Lambda, \Delta^\Lambda, \delta^\Lambda, P^\Lambda, Q^\Lambda$ will eventually have to be expressed in terms of the physical parameters of a given solution (mass, angular momentum, NUT charge, electric and magnetic charges) upon solving the equations of motion in a chosen model. The question whether the above ansatz leads to solutions in models of the STU type is left for future research.

\acknowledgments

We would like to thank S.\ Katmadas, A.\ Tomasiello and S.\ Vandoren for interesting discussions. A.G.\ and C.T.\ would also like to thank Università degli Studi di Milano and Milano-Bicocca   for the kind hospitality during various visits, while this work was in progress. 

A.G.\ and C.T.\ acknowledge support by the Netherlands Organization for Scientific Research (NWO) under the VICI grant 680-47-603. K.H.\ is supported in part by INFN and by the MIUR-FIRB grant RBFR10QS5J ``String Theory and Fundamental Interactions''. The work of D.K.~was partially supported by INFN and MIUR-PRIN contract 2009-KHZKRX. The work of O.V.\ is supported by the German Science Foundation (DFG) under the Collaborative Research Center (SFB) 676 ``Particles, Strings and the Early Universe''. 

\appendix

\section{1/2 BPS near-horizon geometries}
\label{1/2BPS}

An interesting class of half-supersymmetric backgrounds was obtained in \cite{Klemm:2010mc}.
It includes the near-horizon geometry of extremal rotating black holes. The metric and the
fluxes read respectively
\begin{eqnarray}
ds^2 &=& -z^2e^{\xi}\left[dt+4(e^{-2\xi}-L)\frac{dx}z\right]^2 + 4e^{-\xi}\frac{dz^2}{z^2}
\nonumber \\
&& \qquad +16e^{-\xi}(e^{-2\xi}-L)dx^2 + \frac{4e^{-2\xi}d\xi^2}{Y^2(e^{-\xi}-Le^{\xi})}\ ,
\label{metr-Y}
\end{eqnarray}
\begin{eqnarray}
F^\Lambda&=&8i\left(\frac{\bar X L^\Lambda}{1-iY}-\frac{X \bar L^\Lambda}{1+iY}\right)dt\wedge dz
\label{fluxes-half-BPS} \\
&&+\frac 4Y\left[\frac{2\bar X L^\Lambda}{1-iY}+\frac{2X\bar L^\Lambda}{1+iY}+\left(\mbox{Im}\,\mathcal{N}
\right)^{-1|\Lambda\Sigma}g_\Sigma\right](zdt-4Ldx)\wedge d\xi\ , \nonumber
\end{eqnarray}
where $L$ is a real integration constant, $X\equiv g_\Lambda L^\Lambda$, and $Y$ is defined by
\eq
Y^2=64e^{-\xi}|X|^2-1\ . \label{Y}
\feq
The moduli fields $z^{\alpha}$ depend on the coordinate $\xi$ only, and obey the differential equation
\eq
\frac{dz^{\alpha}}{d\xi} = \frac i{2\bar X Y}(1-iY)g^{\alpha\bar\beta}{\cal D}_{\bar\beta}\bar X\ .
\label{dzdxi}
\feq
For $L>0$, the line element \eqref{metr-Y} can be cast into the simple form
\begin{eqnarray}
ds^2 &=& 4e^{-\xi}\left(-z^2d{\hat t}^2 + \frac{dz^2}{z^2}\right) + 16L(e^{-\xi}-Le^{\xi})
\left(dx - \frac z{2\sqrt L}d\hat t\right)^2 \nonumber \\
&& \qquad + \frac{4e^{-2\xi}d\xi^2}{Y^2(e^{-\xi}-Le^{\xi})}\ , \label{near-hor}
\end{eqnarray}
where $\hat t\equiv t/(2\sqrt L)$. \eqref{near-hor} is of the form (3.3) of
\cite{Astefanesei:2006dd}, and describes the near-horizon geometry of extremal
rotating black holes\footnote{Metrics of the type \eqref{near-hor} were discussed for the first
time in \cite{Bardeen:1999px} in the context of the extremal Kerr throat geometry.},
with isometry group $\text{SL}(2,\bR)\times\text{U}(1)$.
From \eqref{dzdxi} it is clear that the scalar fields have
a nontrivial dependence on the horizon coordinate $\xi$ unless $g_\Lambda{\cal D}_{\alpha}X^\Lambda=0$.
As was shown in \cite{Klemm:2010mc}, the solution with constant scalars is the near-horizon
limit of the supersymmetric rotating hyperbolic black holes in minimal gauged
supergravity \cite{Caldarelli:1998hg}. Moreover, in \cite{Klemm:2011xw}, a class of rotating BPS
black holes with running scalar was constructed for the prepotential $F=-iX^0X^1$, which has
again \eqref{near-hor} as near-horizon limit.

Let us first consider the case of gauged supergravity with flat scalar potential $V$, that was studied recently
in \cite{Hristov:2012nu}. The condition $V=0$ translates into
\eq
{\cal D}_{\alpha}Xg^{\alpha\bar\beta}{\cal D}_{\bar\beta}\bar X = 3|X|^2\,. \label{flat-pot}
\feq
Now, using \eqref{dzdxi}, it is straightforward to show that
\eq
\frac{d|X|^2}{d\xi} = {\cal D}_{\alpha}Xg^{\alpha\bar\beta}{\cal D}_{\bar\beta}\bar X\,.
\feq
Using \eqref{flat-pot}, this can be integrated to give
\eq
|X|^2 = C^2e^{3\xi}\,,
\feq
where $C$ denotes an integration constant. Then, \eqref{Y} allows to express $\xi$ in terms of $Y$,
\eq
e^{-\xi} = 8C(Y^2 + 1)^{-1/2}\,. \label{xi-Y}
\feq
In asymptotically flat space, there are underrotating black holes \cite{Rasheed:1995zv}, whose
near-horizon geometry is given by
\eq
ds^2 = -e^{2U}r^2(dt + \omega)^2 + e^{-2U}\left(\frac{dr^2}{r^2} + d\theta^2 + \sin^2\theta d\phi^2\right)\,,
\label{near-hor-underrot}
\feq
where
\eq
e^{-4U} = -I_4 - j^2\cos^2\theta\,, \qquad \omega = j\frac{\sin^2\theta}{r}d\phi\,,
\feq
$j$ is the rotation parameter and $I_4$ denotes the quartic invariant of the charges.
It turns out that \eqref{near-hor-underrot} is actually a one-half BPS solution
of gauged supergravity (with flat potential). To see this, choose $C=|I_4|^{1/2}/32$ (we need $I_4<0$),
$L=(|I_4|-j^2)/16$, and make the coordinate transformation
\eq
z = \frac r2\,, \qquad x=\frac{2\phi}j\,, \qquad e^{-\xi} = \frac14 e^{-2U}\,,
\feq
that casts \eqref{metr-Y} into \eqref{near-hor-underrot}.
(Use also \eqref{xi-Y} to eliminate $Y$ in favour of $\xi$). In \cite{Hristov:2012nu} it was shown that
the underrotating near-horizon geometry \eqref{near-hor-underrot} preserves at least one quarter
of the supersymmetries, but it was not excluded that it is even 1/2 BPS (cf.~footnote 14 of
\cite{Hristov:2012nu}). Here we showed that this is indeed the case.

In the case of the model with prepotential $F=-iX^0X^1$, the general solution of the
differential equation \eqref{dzdxi} was found in \cite{Klemm:2011xw}, and is given by
\eq
\tau = \frac{X^1}{X^0} = \frac{g_0}{g_1}\frac{Y-i+C}{Y-i-C}\,, \label{tau-half-BPS-X0X1}
\feq
where $C$ denotes a complex integration constant. We shall now obtain the conditions under which
the near-horizon geometry of the general solution of section \ref{nut} fits into this 1/2 BPS class.
In the extremal limit, the function $Q$ has a double root at $q=q_{\text h}$. Defining new coordinates
$z,\hat t,\hat y$ by
\begin{displaymath}
q = q_{\text h} + \epsilon q_0 z\,, \quad t = \frac{\hat t q_0}{\epsilon}\,, \quad y = \hat y
-\frac{\hat t q_0}{(q_{\text h}^2 - \Delta^2)\epsilon}\,, \quad q_0^2 \equiv \frac{q_{\text h}^2 -
\Delta^2}{\hat C}\,, \quad \hat C \equiv a_2 + \frac{6q_{\text h}^2}{l^2}\,,
\end{displaymath}
and taking $\epsilon\to 0$ with $z,\hat t,\hat y$ fixed, we get the near-horizon limit of
\eqref{ansatz-nonextr-el}, \eqref{tau-NUT} and the field strength following from \eqref{Ai},
\eq
ds^2 = \frac{W_{\text h}}{\hat C}\left(-z^2 d\hat t^2 + \frac{dz^2}{z^2}\right) + \frac{W_{\text h}dp^2}{P}
            + \frac{P}{W_{\text h}}\left((q_{\text h}^2 - \Delta^2) d\hat y - \frac{2q_{\text h}z}{\hat C}
            d\hat t\right)^2\,,
\feq
\eq
\tau = \frac{g_0}{g_1}\frac{q_{\text h} + \Delta - i(p + \delta)}{q_{\text h} - \Delta - i(p - \delta)}\,,
\feq
\begin{eqnarray}
F^{\Lambda} &=& \frac{\mathsf{P}^{\Lambda}((p - \delta)^2 - q_{\text h}^2 + \Delta^2) +
2p\mathsf{Q}^{\Lambda}(q_{\text h} - \Delta)}{W_{\text h}^2}\left[(q_{\text h}^2 - \Delta^2) d\hat y
- \frac{2q_{\text h}z}{\hat C}d\hat t\right]\wedge dp \nonumber \\
&&\quad + \frac{\mathsf{Q}^{\Lambda}((q_{\text h} - \Delta)^2 - p^2 + \delta^2) + 2q_{\text h}
\mathsf{P}^{\Lambda}( p - \delta)}{W_{\text h}\hat C}d\hat t\wedge dz\,,
\end{eqnarray}
where $W_{\text h}\equiv q^2_{\text h}-\Delta^2+p^2-\delta^2$.
This coincides with \eqref{near-hor}, \eqref{tau-half-BPS-X0X1} and \eqref{fluxes-half-BPS}
respectively if the following constraints hold:
\eq
Q = \frac1{l^2}(q^2 - q_{\text h}^2)^2\,, \qquad \mathsf{Q}^{\Lambda} = \delta = 0\,, \qquad
\mathsf{P}^{\Lambda} = \frac{8q_{\text h}^2}{l^4 g_{\Lambda}}\sqrt L\,.
\feq
The coordinates $\hat y$ and $p$ are related to $x$ and $\xi$ in \eqref{near-hor} by
\eq
\hat y = \frac{l^2\sqrt L}{q_{\text h}(q_{\text h}^2 - \Delta^2)}x\,, \qquad e^{\xi} =
\frac{16 q_{\text h}^2}{l^2 W_{\text h}}\,,
\feq
while the constant $C$ in \eqref{tau-half-BPS-X0X1} is given by $C=-i\Delta/q_{\text h}$,
and $Y=-p/q_{\text h}$. Since the location of the horizon can be set to $q_{\text h}=l$ without loss
of generality by using the scaling symmetry \eqref{scaling-symm}, the 1/2 BPS near-horizon geometry
is specified in terms of the two parameters $\Delta$ (or alternatively $C$) and $L$. Note that this
solution was first constructed in \cite{Klemm:2011xw}. Since \eqref{fluxes-half-BPS}, \eqref{near-hor}
and \eqref{tau-half-BPS-X0X1} do not exhaust all possible half-supersymmetric solutions,
the results of this appendix do not necessarily imply that there are no further 1/2 BPS subclasses of
\eqref{ansatz-nonextr-el}-\eqref{tau-NUT}.

\section{Real formulation of special geometry}
\label{AppB}

Here we will present three dimensional effective Lagrangian for stationary field configurations, adapted to the real formulation of special geometry.
In the ungauged case, the resulting three-dimensional Lagrangian, which takes the form of a non-linear sigma model, was constructed in \cite{Mohaupt:2011aa} (see eq (29)). For the case of gauged supergravity one must add an additional Fayet-Iliopoulos potential 
\begin{eqnarray}
 	\label{eq:AppB3dLag} 
	\begin{aligned}
			\texttt{e}^{-1}_3 {\cal L}_3 = \;\;&\frac{1}{2} R_3 - \tilde{H}_{ab} \left(\partial q^a \partial q^b - \partial \hat{q}^a \partial \hat{q}^b \right) + \frac{1}{2H} V(q) \\ 
			&- \frac{1}{H^2} \left( q^a \Omega_{ab} \partial q^b \right)^2 + \frac{2}{H^2} \left( q^a \Omega_{ab} \partial \hat{q}^b \right)^2  \\
			&- \frac{1}{4 H^2} \left( \partial \tilde{\phi} + 2\hat{q}^a \Omega_{ab} \partial \hat{q}^b  \right)^2 \;,
 	\end{aligned}
\end{eqnarray}
where $\tilde{H}_{ab} = -\frac12 \frac{\partial^2}{\partial q^a \partial q^b}  \log( -2 H )$. The Lagrangian is completely determined by specifying a Hesse potential $H$, which plays the an analogous role to the holomorphic prepotential when using special real coordinates. 
In \cite{Klemm:2012yg, Klemm:2012vm} static solutions were considered for which the second and third lines of the above Lagrangian can be consistently set to zero. In this paper we are interested in NUT charged and rotating solutions in which case all terms become relevant.

The fields appearing in (\ref{eq:AppB3dLag}) are related to the complex scalar fields and gauge fields appearing in the main text by the following dictionary:
\begin{align}
	q^a &:= \left( \begin{array}{c} x^\Lambda \\ y_\Lambda \end{array} \right)  := \left( \begin{array}{c} e^{\phi/2} \text{Re} L^\Lambda \\ e^{\phi/2} \text{Re} M_\Lambda \end{array} \right) \;, \label{eq:AppBdic1} \\
	\partial_m \hat{q}^a &:= \left( \begin{array}{c} \frac12 \partial_m \zeta^\Lambda \\ \frac12 \partial_m \tilde{\zeta}_\Lambda \end{array} \right)  :=\left( \begin{array}{c} \frac12 F^\Lambda_{m 0} \\ \frac12 G_{\Lambda|m 0}\end{array} \right) \;, \label{eq:AppBdic2}\\
	\left( \partial_m \tilde{\phi} + 2 \hat{q}^a \Omega_{ab} \partial_m \hat{q}^b\right) &:= 2H^2 \varepsilon_m^{\;\;\;n r} \partial_{[n} \omega_{r]}\;, \label{eq:AppBdic3}
\end{align}
whilst the three- and four-dimensional metrics satisfy the expression
\begin{equation}
	g_4 = -e^\phi(dt + \omega_m dx^m)^2 + e^{-\phi} g_3 \;. \label{eq:AppBmetrics}
\end{equation}
The Hesse potential is related to the Fayet--Iliopoulos potential $V$ through expression (A.2) of \cite{Klemm:2012yg}, and to the KK-scalar through $-2H = e^\phi$.

\subsection{Equations of motion}

For future reference it will be convenient to write down the full equations of motion of the three-dimensional effective Lagrangian (\ref{eq:AppB3dLag}).

We first perform the variation with respect to the $q^a$ fields, which results in the equations
\begin{eqnarray}
	\begin{aligned}
			&2\nabla^m \left[\tilde{H}_{ab}\partial_m q^b \right] -  \partial_a \tilde{H}_{bc} \left(\partial_m q^b \partial^m q^c - \partial_m \hat{q}^b \partial^m \hat{q}^c \right) + \partial_a \left( \frac{1}{2H} V(q) \right) \\ 
			&+ 2\nabla^m \left[ \frac{1}{H^2} q^c \Omega_{ca} \left( q^d \Omega_{de} \partial_m q^e \right) \right] \\
			&- 2\partial_a \left( \frac{1}{H} q^c \right) \left[ \Omega_{cb} \partial_m q^b \frac{1}{H} \left( q^d \Omega_{de} \partial^m q^e \right) - 2 \Omega_{cb} \partial_m \hat{q}^b \frac{1}{H} \left( q^d \Omega_{de} \partial^m \hat{q}^e \right) \right] \\
			&- \partial_a \left(\frac{1}{4H^2}\right)\left( \partial_m \tilde{\phi} + 2\hat{q}^c \Omega_{cd} \partial_m \hat{q}^d  \right)\left( \partial^m \tilde{\phi} + 2\hat{q}^c \Omega_{cd} \partial^m \hat{q}^d  \right) = 0\;. 
	\end{aligned}
	\label{eq:AppBeom1} 
\end{eqnarray}
Next, by varying the $\hat{q}^a$ fields we get
\begin{eqnarray}
	\begin{aligned}
			&2 \nabla^m \left[\tilde{H}_{ab}\partial_m \hat{q}^b \right] \\ 
			&+ 4 \nabla^m \left[ \frac{1}{H^2} q^c \Omega_{ca} \left( q^d \Omega_{de} \partial_m \hat{q}^e \right) \right] - \nabla^m \left[ \frac{1}{H^2} \hat{q}^b \Omega_{ba} \left( \partial_m \tilde{\phi} + 2\hat{q}^c \Omega_{cd} \partial_m \hat{q}^d  \right) \right]  \\
			&+ \frac{1}{H^2}\Omega_{ab} \partial_m \hat{q}^b \left( \partial^m \tilde{\phi} + 2\hat{q}^c \Omega_{cd} \partial^m \hat{q}^d  \right) = 0\;.
		\end{aligned}
	\label{eq:AppBeom2}
\end{eqnarray}
	The variation of the $\tilde{\phi}$ field, which descends from the Kaluza--Klein vector, gives us simply
	\begin{equation}
		\nabla^m \left[ \frac{1}{4H^2}\left(\partial_m \tilde{\phi} + 2\hat{q}^c \Omega_{cd} \partial_m \hat{q}^d \right) \right] = 0 \;.
		\label{eq:AppBeom3}
	\end{equation}
	Since the dualisation procedure swaps the role of the field equations and Bianchi identities, this equation gives us simply the  Bianchi identity for the KK-vector. 
Finally, from the variation of the three-dimensional metric we obtain the following Einstein equations
\begin{eqnarray}	
	\begin{aligned}
			&\frac{1}{2} R_{mn} - \tilde{H}_{ab} \left(\partial_m q^a \partial_n q^b - \partial_m \hat{q}^a \partial_n \hat{q}^b \right) + \frac{1}{2H} g_{m n} V(q)   \\ 
			&- \frac{1}{H^2} \left( q^a \Omega_{ab} \partial_m q^b \right)\left( q^c \Omega_{cd} \partial_n q^d \right) + \frac{2}{H^2} \left( q^a \Omega_{ab} \partial_m \hat{q}^b \right)\left( q^c \Omega_{cd} \partial_n \hat{q}^d \right)  \\
			&- \frac{1}{4 H^2} \left( \partial_m \tilde{\phi} + 2\hat{q}^a \Omega_{ab} \partial_m \hat{q}^b \right)\left( \partial_n \tilde{\phi} + 2\hat{q}^c \Omega_{cd} \partial_n \hat{q}^d  \right)  = 0\;. 
	\end{aligned}
	\label{eq:AppBeom4} 
\end{eqnarray}

\subsection{Example: solution of the $F = -i X^0 X^1$ model}

By way of example, we shall present the rotating nonextremal solution of the $F = -i X^0 X^1$ model, as given section 4, in terms of the real formulation of special geometry: 

\[
	g_3 = v\left( \frac{dq^2}{Q} + \frac{dp^2}{P} \right)+ PQ dy^2 \;,
\]

\begin{equation}
	q^a 
	= \frac{v^\frac12}{\mu^\frac12 2W} \left( \begin{array}{c} \vspace{.5em}
	q_1 \\ \vspace{.5em}
	\dfrac{ \mu W q_1 - p_1(\nu p + \lambda q) }{q_1^2 + p_1^2} \\ \vspace{.5em}
	\dfrac{ \mu W p_1 + q_1(\nu p + \lambda q) }{q_1^2 + p_1^2} \\ \vspace{.5em}
	p_1
	\end{array} \right) , 
\;\;\;\;
	\hat{q}^a 
	= \frac{1}{2W}  \left( \begin{array}{c} \vspace{.5em}
	\mathsf{P}^0 p_1 + \mathsf{Q}^0 q_1 \\ \vspace{.5em}
	\mathsf{P}^1 p_1 + \mathsf{Q}^1 q_1 \\ \vspace{.5em}
	\mu (\mathsf{P}^0q_2 - \mathsf{Q}^0p_2) \\ \vspace{.5em}
	\dfrac{1}{\mu}(\mathsf{P}^1q_1 - \mathsf{Q}^1 p_1) \\
	\end{array} \right) , \notag 
\end{equation}
\begin{align}
	\frac{1}{H} &\left( \partial_p \tilde{\phi} + 2 \hat{q}^a \Omega_{ab} \partial_p \hat{q}^b \right)
	= -\frac{v}{PW} \partial_q \omega_y \;, \notag \\ 
	\frac{1}{H} &\left( \partial_q \tilde{\phi} + 2 \hat{q}^a \Omega_{ab} \partial_q \hat{q}^b \right)
	= \frac{v}{QW} \partial_p \omega_y \;. \notag 
\end{align}
Here the functions $Q, P, W,q_1, q_2, p_1, p_2$ and parameters $\mathsf{P}^\Lambda,\mathsf{Q}^\Lambda,\mu, \nu, \lambda$ are those that appear in the solution (\ref{ansatz-nonextr-el}) -- (\ref{eq:nut-mu}), and the KK-vector is given by
\[
	\omega_y = -\frac{(Pq_1q_2 + Qp_1p_2)}{v} \;, \qquad v = Q-P \;.
\]
Using the expression for the Hesse potential (B.1) of \cite{Klemm:2012yg}, one may explicitly check that the above field configuration solves the equations of motion (\ref{eq:AppBeom1}) -- (\ref{eq:AppBeom4}).

\end{document}